\newenvironment{theindex}
\begin{theindex}
\documentclass[journal,twoside,web]{ieeecolor}
\let\labelindent\relax
\usepackage{enumitem}
\usepackage[nottoc]{tocbibind}
\usepackage{siunitx}
\usepackage{subfig}

\usepackage{tmi}
\usepackage{dblfloatfix}
\usepackage{cite}
\usepackage{amsmath,amssymb,amsfonts}
\usepackage{algorithmic}
\usepackage{graphicx}

\usepackage{epstopdf}

\usepackage{textcomp}
\def\BibTeX{{\rm B\kern-.05em{\sc i\kern-.025em b}\kern-.08em
    T\kern-.1667em\lower.7ex\hbox{E}\kern-.125emX}}
\begin{document}



\title{
Deep Learning-Based Sparse Whole-Slide Image Analysis for the Diagnosis of Gastric Intestinal Metaplasia
}

\author{Jon Braatz, Pranav Rajpurkar, Stephanie Zhang, Andrew Y. Ng, Jeanne Shen

\thanks{Correspondence to: J. Braatz, J. Shen}
\thanks{J. Braatz, S. Zhang, and A. Ng are with the Department of Computer Science, Stanford University, USA     (jfbraatz@alumni.stanford.edu, stephaniezhang@stanford.edu, ang@cs.stanford.edu).}
\thanks{
P. Rajpurkar is with the Department of Biomedical Informatics, Harvard University, USA. (pranav\_rajpurkar@hms.harvard.edu)}
\thanks{J. Shen is with the Department of Pathology and the Center for Artificial Intelligence in Medicine \& Imaging, Stanford University, USA (jeannes@stanford.edu)}
}

\maketitle

\begin{abstract} 
In recent years, deep learning has successfully been applied to automate a wide variety of tasks in diagnostic histopathology. However, fast and reliable localization of small-scale regions-of-interest (ROI) has remained a key challenge, as discriminative morphologic features often occupy only a small fraction of a gigapixel-scale whole-slide image (WSI). In this paper, we propose a sparse WSI analysis method for the rapid identification of high-power ROI for WSI-level classification. We develop an evaluation framework inspired by the early classification literature, in order to quantify the tradeoff between diagnostic performance and inference time for sparse analytic approaches. We test our method on a common but time-consuming task in pathology - that of diagnosing gastric intestinal metaplasia (GIM) on hematoxylin and eosin (H\&E)-stained slides from endoscopic biopsy specimens. GIM is a well-known precursor lesion along the pathway to development of gastric cancer. 
We performed a thorough evaluation of the performance and inference time of our approach on a test set of 20 GIM-positive and 86 GIM-negative WSI, finding that our method successfully detects GIM in all positive WSI, with a WSI-level classification area under the receiver operating characteristic curve (AUC) of 0.98 and an average precision (AP) of 0.95. Furthermore, we show that our method can attain these metrics in under one minute on a standard CPU. Our results are applicable toward the goal of developing neural networks that can be easily deployed in clinical settings to support pathologists in quickly localizing and diagnosing small-scale morphologic features in WSI.
\end{abstract}

\begin{IEEEkeywords}
Deep Learning, Convolutional Neural Network, Histopathology, Performance-Efficiency, Intestinal Metaplasia, Gastric Biopsy, Fast Inference, Early Classification
\end{IEEEkeywords}

%
\IEEEpeerreviewmaketitle

\section{Introduction}
\label{sec:introduction}

\IEEEPARstart{I}{n} digital pathology, tissue slides are scanned to generate high-resolution whole-slide images (WSI), which are often gigapixel-scale images with typical sizes of $100,000 \times 100,000$ pixels \cite{dimitriou2019deep}. With the growing adoption of digital pathology, deep neural networks have increasingly been applied to automate a variety of tasks in histopathology \cite{survey}. However, the practical implementation of deep learning for WSI analysis in clinical settings has remained a challenge, due to the high computational requirements typically necessary for keeping inference times within the sub-minute time constraints imposed by clinical workflows \cite{jiang2020emerging,  serag2019translational}.

In this study, we approach the task of quickly and reliably localizing small-scale morphologic features in WSI. The simple presence of some features in a tissue sample (for example, specific cells, glands, or small lesions), may be sufficient to render a particular diagnosis, and determining whether such features are present within a WSI may often be likened to "searching for a needle in a haystack'', due to their small size and scarcity within a gigapixel-scale WSI. Prior research has successfully applied deep learning to such small-scale ROI detection tasks, such as the diagnosis of micro-metastatic disease \cite{steiner2018impact,lin2019fast}, microvascular invasion \cite{janowczyk2016deep}, tumor budding in colorectal carcinoma \cite{bokhorst2018automatic}, and malignant and premalignant lesions in Barrett's esophagus \cite{tomita2019attention}. 
However, most deep learning applications have taken a dense approach to WSI analysis, which relies on the availability of specialized hardware to process an entire slide within reasonable time constraints. Detection of small-scale morphologic ROI using the consumer-grade hardware commonly found in pathologist workstations has thus remained a challenge to the deployment of deep learning in clinical practice.

To address this challenge, we present a sparse WSI analytic approach to detecting the presence of small-scale morphologic features in a WSI, which achieves a high slide-level classification performance and low inference time by processing only a fraction of the slide. Performing the analysis locally on consumer-grade hardware entails processing regions of the slide in sequence; thus, our approach sequentially samples fixed-sized image patches from an input WSI, classifies each patch using a convolutional neural network (CNN), aggregates the obtained patch-level scores, and continuously updates the slide-level classification as patches are sampled. This sparse approach enables an accurate classification to be obtained in roughly the time it takes to successfully identify one region containing the feature in question, rather than processing the entire slide. 

Any sequential processing approach entails making several design choices which affect both diagnostic performance and inference speed, including the method by which patches are chosen and processed, as well as the size of each patch to be processed. To inform these design choices, we adapted evaluation methods from the ``early time series classification'' literature to the digital pathology setting, using this framework to quantify and compare the tradeoff between diagnostic performance and inference time for different choices of patch size.

We apply this to the diagnosis of gastric intestinal metaplasia (GIM), a premalignant condition in which the native gastric epithelium is replaced by non-native, intestinal-type epithelium. \cite{jencks2018overview, piazuelo2013gastric}. GIM is highly prevalent within the general population, is detected in nearly 1 of every 4 patients undergoing upper endoscopy, and is associated with chronic gastritis and a 6-fold increase in the risk of gastric cancer \cite{zullo2012follow}. The histologic diagnosis of GIM relies on the identification of goblet cells, which are round to ovoid mucin-secreting cells characterized by nuclear compression by cytoplasmic acid mucin \cite{paull1976histologic,TGH5866}. The identification of only a single goblet cell within a WSI is sufficient to diagnose the patient as GIM positive. Therefore, in this early classification investigation, we use goblet cell detection as our local "evidence localization" task, and slide-level GIM diagnosis as our global classification task. 

Our contributions in this study are summarized as follows:

\begin{enumerate}[label=(\roman*)]
\item
We present an evaluation framework to characterize the tradeoff between a model's WSI-level diagnostic performance and the number of tissue patches in a WSI that are processed during inference, using evaluation methodologies from the ``early time-series classification'' literature, which can be used to select an optimal model from among a set of candidate sequential models. To the best of our knowledge, this is the first such analysis in the digital pathology literature.
\item
We develop a deep learning-based sparse whole-slide image analysis method that performs diagnosis of GIM on H\&E-stained WSI, while presenting evidence for the diagnosis in the form of discriminative regions of interest containing goblet cells, with an evidence localization performance competitive with that of a theoretical oracle model.
\item Using our evaluation framework, we show that our approach achieves high classification performance within the constraints of clinical practice, attaining an average precision of 0.99 in under one minute using a CPU.

\end{enumerate}

\begin{figure}[ht!]
    \includegraphics[width=0.99\columnwidth]{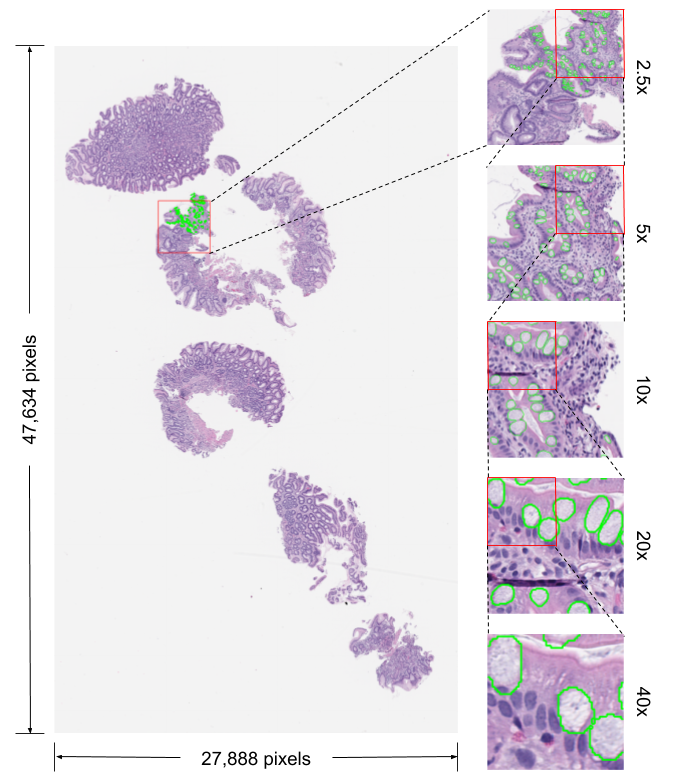}
    \caption{An H\&E-stained whole-slide image (WSI) of a gastric biopsy diagnosed as positive for intestinal metaplasia. Patches from the slide at each magnification are shown, with goblet cells segmented in green. }
    \label{fig:slide_annotations}
\end{figure}
\section{Related Work}

Some progress towards reducing inference time using sparse analysis of input signals has been made independently in the areas of digital pathology and early time series classification. 

\subsection{Sparse Analysis for Digital Pathology}
A prominent example of sparse analysis in digital pathology is HASHI \cite{cruz2018high}, which takes a Quasi Monte-Carlo patch-sampling approach to output a dense probability map for invasive breast cancer detection. As the output is a probability map for breast cancer, the average Dice coefficient is measured as a function of patches sampled, for various sampling strategies. In that study, the adaptive random sampling strategies outperformed a regular in-order tiling strategy. Similarly, in the present study, we measure classification performance as a function of patches sampled for a single, uniformly random strategy, and compare this tradeoff across patch magnifications and training set sizes, rather than sampling strategy.

Huang et al. \cite{huang2011time} attempted to address the problem of time-efficient determination of the nuclear pleomorphism score in breast cancer WSI. Their study used a support vector machine (SVM) to identify invasive and non-invasive cancer patches across WSI at a low magnification, followed by a second step in which ROI from regions with a higher nuclear pleomorphism score were selected using a dynamic sampling method based on Voronoi tessellation, with calculation of a final nuclear pleomorphism score from higher-scaled versions of the selected ROI. In their approach, only 3.3\% of high-magnification patches were thoroughly analyzed, leading, on average, to faster inference times, but reduced performance. In the present work, our method generates outputs continuously and has a high probability of achieving high performance metrics upon processing a fraction of the slide, without having to restrict the analysis to a portion of the slide.

\subsection{Early Classification}

In parallel with these efforts in digital pathology, work in the field of time-series analysis has led to the development of frameworks for fast inference, which classify an input signal without processing it in its entirety, as well as the creation of evaluation frameworks for quantifying how binary classification metrics depend on the fraction of the signal processed \cite{santos2016literature, xing2012early}. The objective in early time series classification is to classify a time series with as few temporal observations as possible, while minimizing the classification accuracy loss. Dennis et al. \cite{dennis2018multiple} studied the problem of fast and efficient classification of sequential data on tiny devices, which is critical for various IoT-related applications such as audio keyword detection or gesture detection. Such tasks are cast as a standard classification task by sliding windows over the datastream to construct data points, analogous to the sliding window approach often taken for WSI analysis. As in our work, the authors cite the resource constraint of inference hardware as a key challenge which they address, and they identify two characteristics of input signals which make their task amenable to early classification. Specifically, they note that, for the time series under consideration, (a) the "signature" of a particular class (e.g. an audio keyword) typically occupies a small fraction of the overall data, and (b) class signatures tend to be discernible early on in the data. Their proposed method, EMI-RNN, uses multiple instance learning (MIL) and an early prediction technique to exploit these observations and increase accuracy over baseline models, while decreasing inference time. 

Sequential processing of patches in WSI can likewise be considered a time series processing task, with the time series consisting of the sequence of patches being processed. Diagnosis of GIM shares the two characteristics identified by Dennis et al, with the "signature" being patches containing goblet cells, which are discernible early on in the data with suitable patch ordering choices. Indeed, these characteristics are shared by any diagnostic task that is amenable to MIL, as these tasks assume that the presence of a diagnostic signature is sufficient to make a classification.

Early classification tasks typically measure performance by quantifying the tradeoff between a binary classification metric, such as accuracy, and the fraction of the input signal processed before making a classification. The time series in these studies are often of the same length, and thus, the fraction of the input signal is a standardized measure across instances in the test set. Whole-slide image processing differs in this respect, as the variance in the size of a slide can be significant. We therefore apply analogous evaluation methods from the time series literature, but replace the fraction of the input signal with the number of patches processed, as our measure of cost.

\section{Data}
\subsection{Whole Slide Image Dataset}
A total of 318 H\&E-stained WSI (60 IM positive and 258 IM negative) from 251 unique patients undergoing upper endoscopy with gastric biopsy at Stanford University Medical Center between January 1, 2015 and December 31, 2018, were digitized at 40x objective magnification (\SI{0.2524 }{\micro\metre} per pixel) using a Leica AT2 scanner (Leica Biosystems, Nussloch, Germany). 51 patients (32 female, 25 male) had at least one GIM-positive WSI, with a mean patient age of 61.7 years (SD 12.3), and 194 patients (108 female, 86 male) had all GIM-negative slides, with a mean patient age of 50.3 years (SD 20.8). The presence of GIM was confirmed by re-review of all WSI by a U.S. board-certified, GI fellowship-trained subspecialty pathologist. There were no diagnostic discrepancies between the diagnosis rendered at the time of original review (based on the pathology report) and that of the reference pathologist.

Some WSI contained up to five serial sections cut from the same tissue block arranged in a row. For these cases, the WSI was cropped to retain only the left-most serial section. WSI-level labels consisted of the GIM status (positive or negative), and pathologist-drawn segmentation maps indicating the locations of goblet cells on all GIM-positive slides, using the open-source QuPath WSI-annotation software \cite{bankhead2017qupath}. To reduce disk space usage and processing time without degrading the quality of the annotations, segmentation maps were downsampled by a factor of four and exported as PNG files. From these PNG files, we derived summed area tables, or integral images \cite{ehsan2015integral}, to support fast constant-time queries for the total number of pixels within axis-aligned square regions of WSI. The median area of each goblet cell annotation was approximately \SI{35}{\micro\metre^2}, and the median number of annotated goblet cells in GIM-positive slides was 132.

The labeled WSI were randomly partitioned into training, validation and test sets, with each split maintaining a 23\% GIM positive to negative ratio, as present in the overall study sample, with all WSI from the same patient assigned to the same split, to avoid patient overlap between training, validation, and test sets. The training dataset consisted of 170 WSI in total, of which 32 slides were GIM-positive and 138 were GIM-negative. The validation dataset, used to select models and model parameters, consisted of 8 GIM-positive slides and 34 GIM-negative slides. The test dataset consisted of 106 WSI (20 positive, 86 negative), used for final reporting of performance metrics. 

\subsection{Patch Sets}
We tiled each slide in the training, validation, and test sets into square patches, using a sliding window approach at pre-specified magnification levels of 2.5x (\SI{904.6}{\micro\metre}), 5x (\SI{452.3}{\micro\metre}), 10x (\SI{226.2}{\micro\metre}), 20x (\SI{113.1}{\micro\metre}), and 40x (\SI{56.5}{\micro\metre}), with all tiles downsampled to a size of 224x224 pixels from the cropped 40x WSI using Lanczos interpolation (the values denoted in parentheses are the lengths of each 224x224 pixel square tile). 

 We used a magnification-dependent stride when extracting patches to keep the total number of patches at each magnification of comparable size (within a factor of 2 between each magnification), as the 40x patch set would have contained orders of magnitude more patches than the 2.5x patch set if a constant stride had been used. The patches in each 40x pool consisted of non-overlapping $224 \times 224$ pixel tiles (i.e. a stride of 224 pixels in the horizontal and vertical directions) and, in general, the stride for patches at magnification $m$ was $224 \cdot \frac{m}{40}$. Non-tissue white background was discarded by retaining only patches containing at least 20\% tissue, with the tissue quantity in each patch determined using tissue region detection via a Gaussian blur operation with a $\sigma$ value of 2.5 on an inverted grayscale thumbnail of the slide at 2x magnification, followed by Otsu thresholding \cite{otsu1979threshold}. Tissue pixels were assigned a pixel value of 1, with all others assigned a pixel value of 0, yielding a binary mask from which the quantity of tissue in each patch was determined.

Each patch was assigned a binary label of positive or negative for GIM (with positive patches defined as those containing at least one goblet cell in their goblet cell segmentation map). All patches from GIM-negative WSI were automatically assigned a negative label, while labels for patches from GIM-positive WSI were determined using the goblet cell segmentation maps. For each patch within a GIM-positive WSI, the summed area table of the segmentation map was used to obtain the number of goblet cell pixels within the patch, with the binary patch-level label assigned based on whether this quantity was above or below a threshold value corresponding to the size of a single goblet cell (chosen as the $10^{th}$ percentile in the size distribution of all goblet cell annotations within all segmentation maps from all WSI).

\section{Methods}
\subsection{Sparse WSI Analysis for Evidence Localization and Slide Classification}
Our tasks for a given input slide are to search for a region containing at least one goblet cell, as evidence for a  positive GIM diagnosis (“evidence localization”), and to output a GIM classification for the slide (“slide classification”). The number of patches processed before finding a goblet cell is dependent on the size of each patch and the order in which patches are processed. We chose a uniformly random patch ordering, as processing patches in a deterministic order could have led to high inference times, depending on the location of goblet cells within a slide. With this randomized strategy, the expected proportion of total patches sampled before encountering a positive patch is roughly equal to the proportion of patches which are positive for goblet cells.

The performance of a CNN trained to recognize goblet cells within a patch is also dependent on the size of the patch. In particular, smaller patch sizes require more patches to be processed before identifying a positive one. Smaller patch sizes also enable higher magnifications to be used, so that goblet cells appear larger within a patch, enabling a CNN to more effectively identify them. The patch size is therefore a critical parameter affecting the tradeoff between performance and inference time. 

We developed a framework that can quantify the tradeoff between performance and inference time in a manner agnostic to both the morphological features being localized and the machine learning approach used to classify whether they are present in a given patch. Using our framework, a patch size can be chosen which maximizes classification performance while keeping slide inference times within clinical time constraints. As the creation of pixel-level annotations of GIM-positive slides is a very time-consuming task for pathologists, we also investigated the relationship between the quantity of annotated GIM-positive WSI used for model development and resultant model performance. We tested our framework for patch magnifications of 2.5x, 5x, 10x, 20x, and 40x, with the size of each field of view provided (in micrometers) for each magnification in Section IIA.

\subsection{Inference Strategy}
For both the evidence localization and slide classification tasks, we first tile the slide without overlap at a chosen patch magnification. We downsample the patches to $224 \times 224$ pixels and reject non-tissue tiles using the approach described in the patch dataset creation step (see Section IIB). We then randomly sample from these patches without replacement and use a patch classification model appropriate to that magnification to sequentially obtain patch scores for each patch.

Our approach to evidence localization is to maintain a ``candidate list'' consisting of the top three patches with the highest scores output by the goblet cell detection model out of all patches sampled thus far. We perform slide classification by aggregating the patch scores from the candidate list into a slide score representing the probability that the slide was GIM-positive. Aggregation is typically performed using a maximum or averaging operation over some or all of the patch-level predictions \cite{coudray_classification_2018, dimitriou2019deep, kraus2016classifying}. In our experiments, the slide classification at each time step consisted of the average of the model scores for patches on the candidate list. The candidate list and slide classification score were updated each time a new patch was sampled, enabling the model to provide continuous outputs for each task as the slide was processed. Within a clinical setting, the candidate list containing the top three patches might also be useful for continuously displaying to a pathologist the patches that are most likely to contain goblet cells as the slide is processed. Our experiments quantified both the number of patch samples required for a positive patch to enter the candidate list for the evidence localization task, and the classification performance as a function of the number of patch samples processed for the slide classification task.

\subsection{Models and Training}
 
Several CNNs were trained separately on fixed-magnification patch sets to classify WSI patches as positive or negative for goblet cells, using the labeled patch sets extracted from the training and validation set slides. We used a model architecture from the EfficientNet family, which has achieved superior accuracy and efficiency on natural image classification by using neural architecture search to design a new baseline network, which was then scaled up \cite{tan_efficientnet_2019}. Specifically, we used EfficientNet-B1, which is 7.6x smaller and 5.7x faster than ResNet-152 \cite{he2016deep}. We trained one model for each of five magnification-specific patch datasets, initializing each model using pretrained ImageNet weights and preprocessing each patch by mean-centering with respect to the ImageNet dataset \cite{deng2009imagenet}.

During each training iteration, half of the examples were randomly sampled from among the positively-labeled patches, with the other half sampled from the negatively-labeled patches. Unweighted binary cross-entropy was used as the loss function. We utilized the Imbalanced-learn library \cite{JMLR:v18:16-365} to randomly sample an equal number of tiles from the majority (negative) class, with replacement. We used the Adam optimizer \cite{kingma2014adam} with a learning rate of $.0001$ and a batch size of 32. The AUC on the corresponding validation patch set was monitored during training, and the training procedure was stopped once this metric did not increase for 100 consecutive iterations. Data augmentation was applied during training, rotating each tile by a multiple of 90 degrees chosen uniformly at random and performing a reflection across the vertical axis with a probability of $1/2$.

As creating goblet cell segmentation maps for GIM-positive slides is manually intensive and requires the full slide to be annotated for accurate patch labels to be obtained, we investigate how the quantity of annotated slides in the training set affected patch classification performance by randomly stratifying the training and validation sets into four ``supervision groups'', with varying numbers of GIM-positive slides used in training and validation, and obtaining performance metrics for each group. The groups consisted of 8 and 2, 16 and 4, 24 and 6, and 32 and 8 annotated GIM-positive slides in the training and validation sets, respectively. Each supervision group contained all patches from all available GIM-negative slides in the training and validation sets, as these patch labels can be concluded to all be negative and hence are readily obtained. We trained 20 models in total, one for each combination of patch magnification and supervision group.

\section{Experiments}
The tasks of evidence localization and slide-level classification, while related, are qualitatively different, as ROI extraction is fundamentally an information retrieval (IR) task and slide-level classification is a binary classification task, in the case of GIM diagnosis. Performance on these tasks depends heavily on a model's patch classification. For this reason, we performed three sets of experiments, with the first evaluating each model's patch classification performance across all patches in the test set, and the latter two sets of experiments evaluating performance and inference times for each of the two tasks. All experiments were conducted using the test set of 20 GIM-positive slides and 86 GIM-negative slides. 

\subsection{Patch Classification}
\subsubsection{Metrics}
Our inference strategy for processing an entire slide, regardless of the chosen patch magnification, required assigning a score to each patch representing the likelihood that the patch contained goblet cells. As the distribution of positive and negative patches in slides is typically heavily imbalanced, with a  majority of negative patches, we first measured a standard metric used in IR and object detection for this setting, namely the average precision (AP). We calculated patch classification APs for each combination of magnification and supervision group, across all patches in the test set at the corresponding magnification.
\subsubsection{Results}
\begin{figure}[ht!]
    \includegraphics[width=\columnwidth]{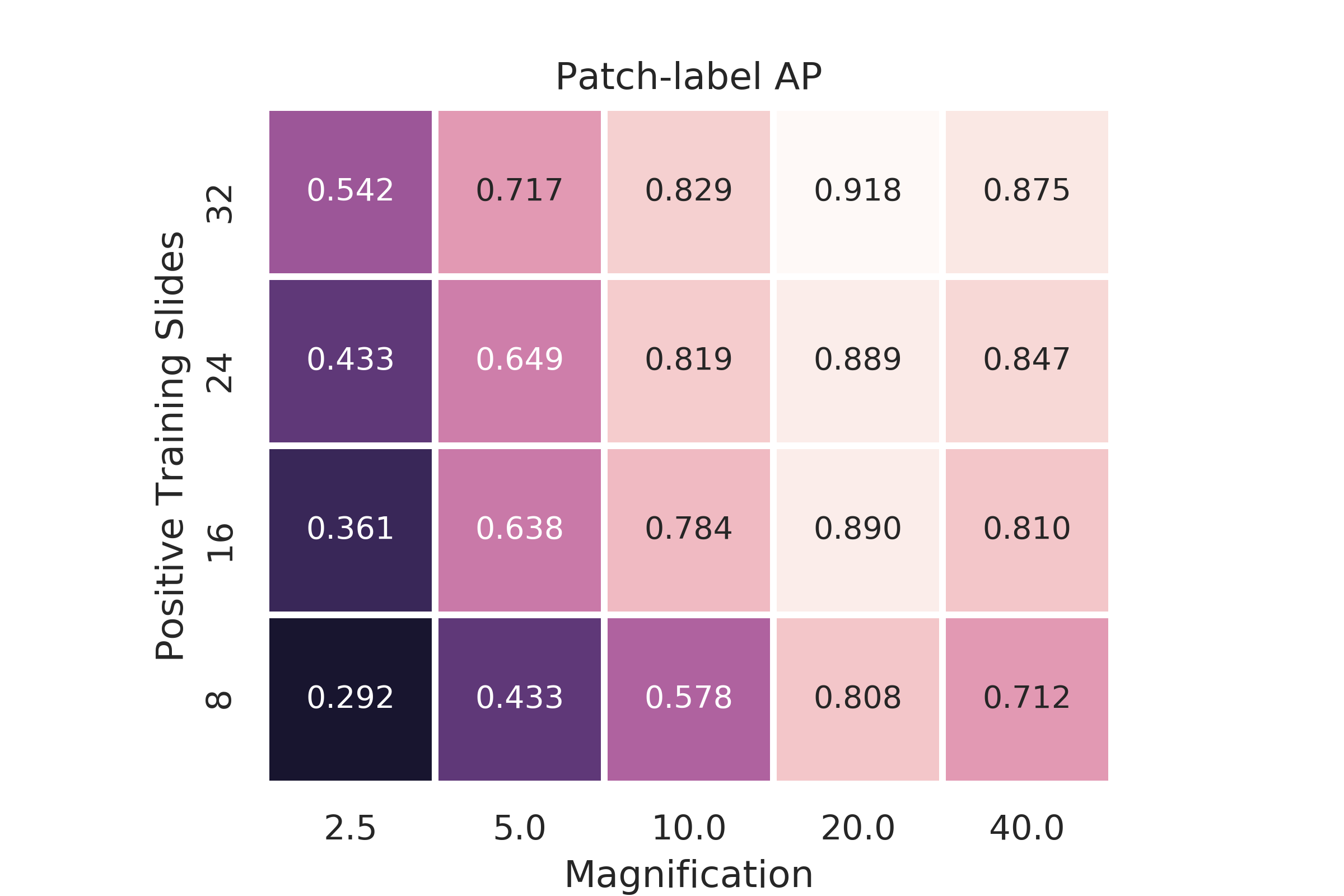}
    \caption{Average precision (AP) of patch-level classification on all GIM-positive slides in the test set, with patches extracted using tiling with no overlap. Classifiers were trained with varying patch magnification levels and  quantities of pixel-level annotations.}
    \label{fig:patch_ap}
\end{figure}
We found that the AP for patch classification increased with the number of positive training examples, reaching a maximum of 0.918 at 20x magnification, when trained using the full training and validation sets (see \ref{fig:patch_ap}). Lower magnification models were found to benefit more substantially from larger training and validation sets than higher magnification models.
The 2.5x model saw an increase in AP from 0.361 to 0.542 when the size of the supervision group was doubled from 16 to 32 GIM-positive slides; the 20x model only saw an increase in AP from 0.890 to 0.918 over this range.
With the full training and validation sets, the 20x model had the highest AP, followed by 40x, 10x, 5x, and 2.5x, in that order. These trends suggest that the increased tissue context from the larger field of view of the 20x compared to the 40x model aids classification performance up to a point, with performance decreasing with decreasing magnification, as the relative size of the goblet cells within a patch decreases.
\subsection{Evidence Localization}
\subsubsection{Metrics}
As patches are sampled at inference time, it is reasonable to expect that, upon a positive patch entering the candidate list, the slide label prediction will not only reflect this, but that, if a pathologist were monitoring the candidate list, they would presumably recognize the patch as being positive and could disregard further model outputs. We therefore considered a positive patch entering the candidate list as the moment of ``success.'' Depending on the order in which patches are processed, it is possible that a success never occurs, so we defined two additional slide-specific metrics, namely the \emph{Failure Rate} and the \emph{Time to Detection}. These metrics capture the probability of failure, as well as the expected time until success, each of which we define with respect to a particular GIM-positive slide. We also define the $\emph{Miss Rate}$ as a summary statistic of the failure rate across the GIM-positive slides in the test set.

For a given slide, we define the \emph{Failure Rate} as the probability that a positive patch never enters the candidate list for a uniformly random permutation of all patches in the slide. This is a function of the highest ranking the model assigns to any positive patch in the slide. For example, if there are 3 negative patches ranked higher than the top-ranked positive patch, then it is possible for these three patches to be processed before any of the positive ones, in which case a positive patch will never enter the candidate list. Letting $r$ denote the top ranking of any positive patch (with the top patch having a rank of 0), $k$ denote the total number of positive patches in the slide, and $T$ denote the size of the candidate list ($T = 3$ in our experiments), the failure rate is given by:
\begin{equation*}
    \frac{\frac{r!}{(r-T)!}}{\frac{(r+k)!}{(r+k-T)!}} = \frac{r!(r+k-T)!}{(r-T)!(r+k)!}
\end{equation*}
when $r \geq T$ and 0 otherwise. If one of the top $T$ patches is positive, then success is guaranteed and the failure rate is 0. We define the \emph{Miss Rate} as the fraction of GIM-positive slides in the test set with a nonzero failure rate.

While not all permutations of the patches may yield a success, we want to quantify how many patches would need to be processed before a success occurs, given that one eventually does. We define the \emph{Time to Detection} as the number of patches processed before a positive patch enters the candidate list, and measure the mean and variance of this quantity over a uniformly random patch permutation. We obtained Monte Carlo estimates of these statistics for each slide over 10,000 trials, and compared these statistics with those from an ``oracle model'' which ranked all positive patches above all negative patches. The mean and variance of the oracle model can be expressed in closed form as the mean and variance of a negative hypergeometric distribution. Concretely, if there are $N$ elements in a set, of which $k$ are successes and $N-k$ are failures, and samples are randomly drawn from the set without replacement, then the number of failures, $t$, until a success is seen, is represented by the negative hypergeometric distribution $NHG_{N, N-k, 1}(t)$, with its mean given by 
\begin{equation*}
    \frac{N-k}{k+1}
\end{equation*}
and variance by 
\begin{equation*}
    \frac{k(N-k)(N+1)}{(k+1)^2(k+2)}.
\end{equation*}
In our case, successes correspond to positive patches and failures correspond to negative patches.

\begingroup
\begin{table}
\caption {\label{tab:table1} Time to Detection} 
\centering
\begin{tabular}{llllll}
& \multicolumn{5}{c}{Patch Magnification}
\cr \cline{2-6}
& \multicolumn{1}{c}{2.5x} & \multicolumn{1}{c}{5x} &
\multicolumn{1}{c}{10x} &
\multicolumn{1}{c}{20x} &
\multicolumn{1}{c}{40x}\\
\hline
Mean (SD) & 2 (1) & 7 (11) & 26 (42) & 60 (117) & 192 (406)\\
Miss Rate & 10\% & 10\% & 0\% & 0\% & 0\%\\
\end{tabular}
\end{table}
\endgroup

\begin{figure}[ht!]
    \includegraphics[width=\columnwidth]{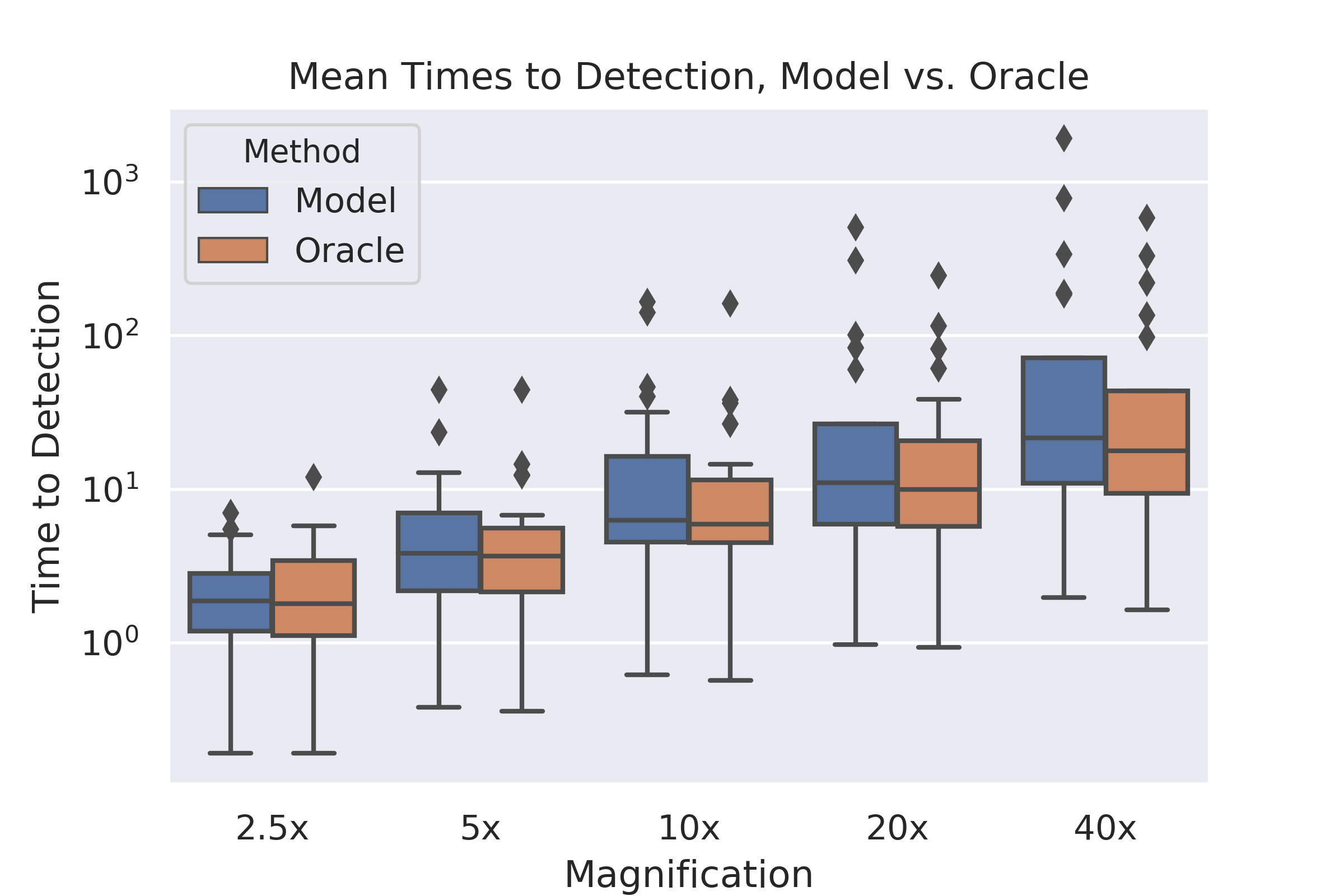}
    \caption{Distribution of mean Time to Detection for evidence localization over GIM-positive test slides at each magnification. Mean Time to Detection reported for each slide is defined as the number of patches processed before a true positive patch enters the candidate list, averaged over 10,000 random permutations of the list of patches for the slide.}
    \label{fig:ttd}
\end{figure}

We found that, at low magnifications such as 2.5x and 5x, we could find positive patches quickly (on the order of dozens of patches sampled), but at the cost of 10\% of slides having a non-zero failure rate. As magnification increased, the failure rates dropped to zero, but at the cost of higher means and standard deviations for the time to detection. As goblet cells only occupy a small fraction of a slide, at higher magnifications such as 20x and 40x, the Time to Detection (TTD) not only increased, but the standard deviation in TTD increased to roughly twice the mean, for the 20x and 40x models. We found that, at a magnification of 10x, we could localize goblet cells quickly while maintaining a miss rate of zero, indicating that patch sizes of 10x afford the shortest time until a positive goblet cell detection, without risking any GIM-positive slides being analyzed without finding a positive patch.

\begin{figure}[ht!]
    \includegraphics[width=\columnwidth]{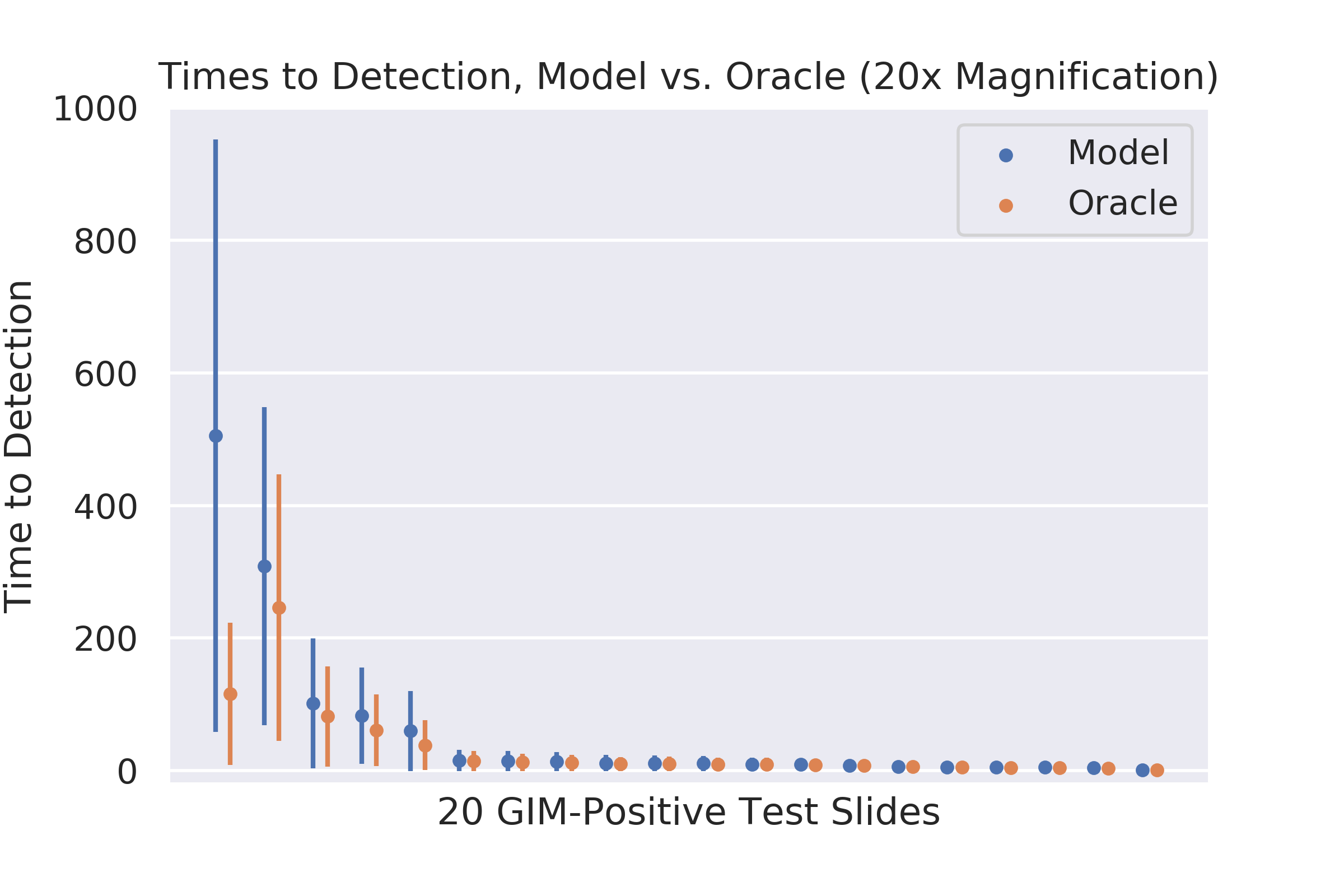}
    \caption{Mean Time to Detection (in seconds) at 20x magnification for positive test slides (represented by a dot), with standard deviations (represented by the line). Our times to detection, determined empirically over 10,000 random permutations of the list of patches for the slide, were compared with those of an oracle, calculated for each slide as the mean and standard deviation of a negative hypergeometric distribution with the number of total tissue patches and positive tissue patches in the slide as parameters.}
    \label{fig:aps_vs_cost_within_size}
\end{figure}

We observed that the median TTDs for all magnifications were relatively short and required fewer than 40 patches to be processed, with mean TTDs competitive with the oracle. The larger values for the mean TTDs were the result of a minority of the test slides having a significantly lower prevalence of goblet cells, necessitating over 100 patches to be processed, on average, for those slides (over 1,000 patches for 40x models, which might be impractical for deployment in a clinical setting). The distribution of mean TTDs for each of the 20 GIM-positive slides in the test set is displayed in Figure \ref{fig:aps_vs_cost_within_size}.

\subsection{Slide Classification}
\subsubsection{Metrics}
The choice of which patch magnification to use depends on the time and performance constraints at inference time. In particular, given the constraints that inference time must not exceed a particular value, and that a slide classification metric must be above a particular performance threshold if it is to be useful in a clinical setting, we would like to choose a patch magnification that achieves the highest possible classification performance.

To inform this choice, we evaluated the slide classification task on the test set by measuring two binary classification metrics, namely the average precision and area under the receiver operating characteristic curve (AUC) of the slide-level predictions, as a function of the number of patches sampled without replacement from a non-overlapping grid of patches in each slide at each magnification. We repeated this 500 times and calculated the means and standards deviations of the two metrics for each magnification. The number of patches sampled was our measure of the cost of the inference strategy, and was proportional to the inference time when patches were processed sequentially. We quantified the tradeoff between classification performance and inference cost for each model, using graphs of mean AP versus the number of patches sampled, for all patch magnifications.  From these, we derived graphs of costs required for each model to reach a desired AP. For a desired value of the AP or AUC, we also obtained estimates of the range of sample sizes required for the metric to reach within one standard deviation of the mean, giving a best and worst case estimate for the inference time required to reach the corresponding level of classification performance.
\subsubsection{Results}
\begin{figure}[ht!]
\centering

\subfloat[]{
  \includegraphics[width=\columnwidth]{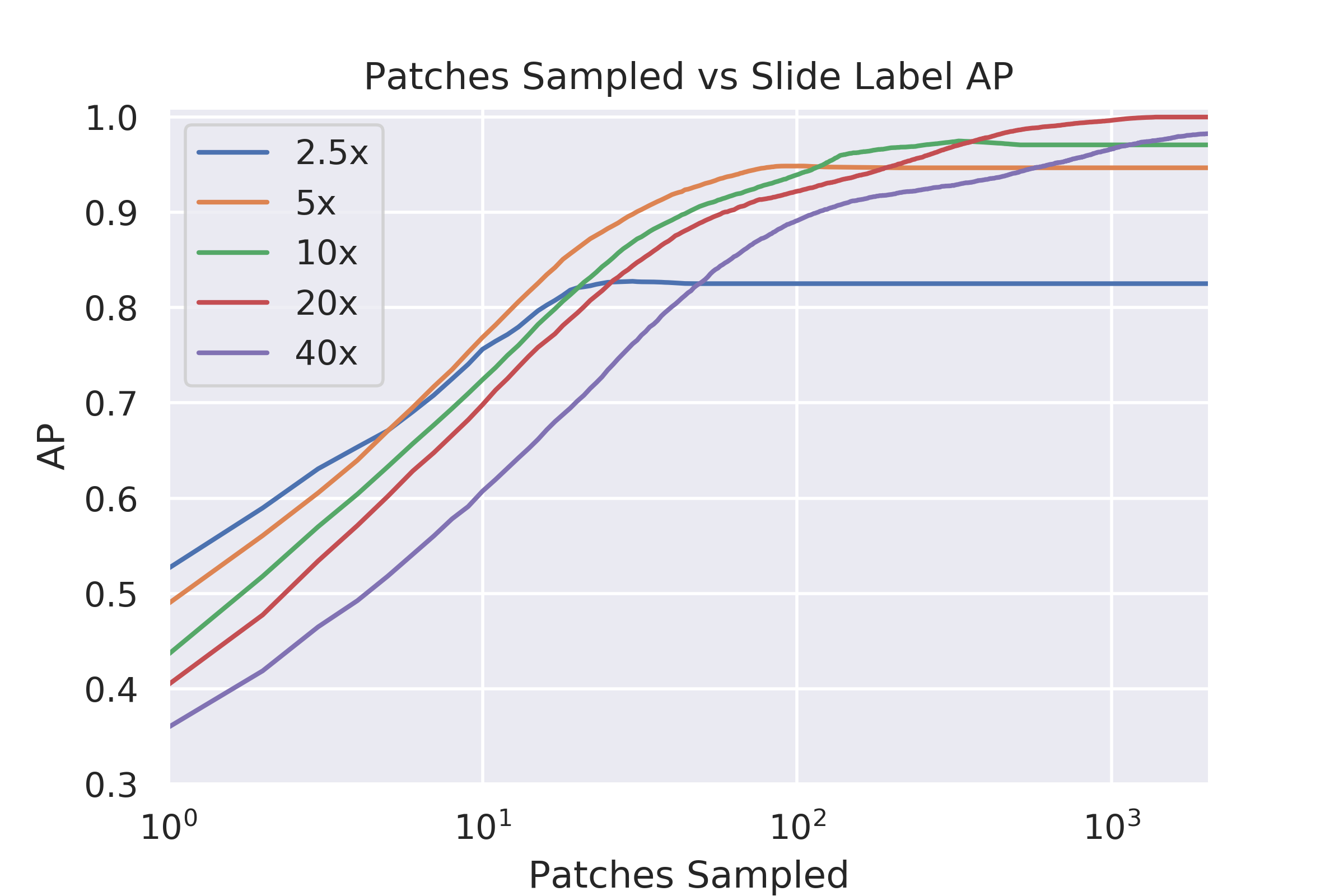}%
}

\subfloat[]{
  \includegraphics[width=\columnwidth]{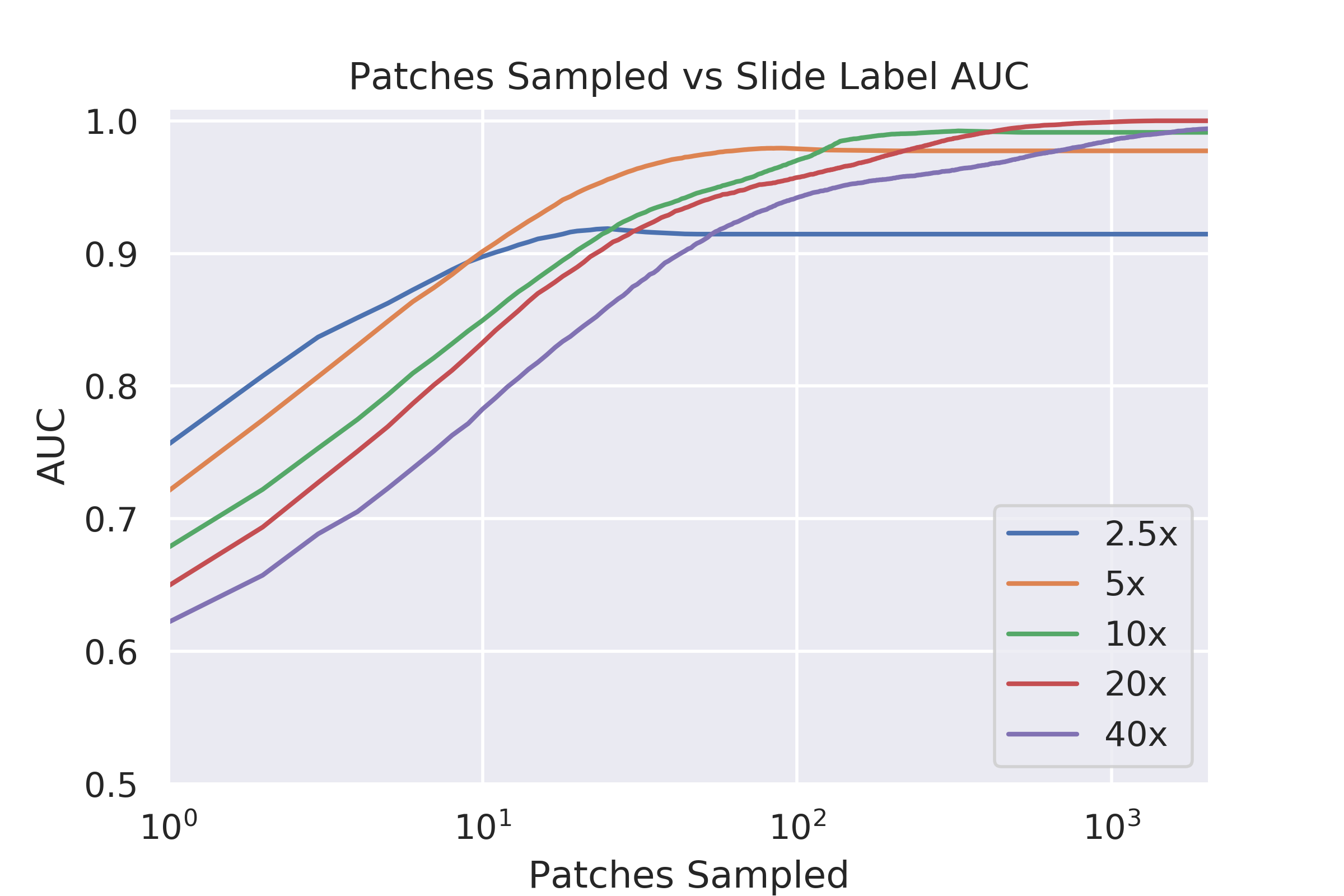}%
}
    
    \caption{ Mean AP (a) and AUC (b), as a function of patches sampled without replacement, of slide-label classifications for all slides in the test set, at all slide magnifications, averaged over 500 replicates.}
    \label{fig:cost_vs_AP}
\end{figure}

\begin{figure}[ht!]
    \includegraphics[width=\columnwidth]{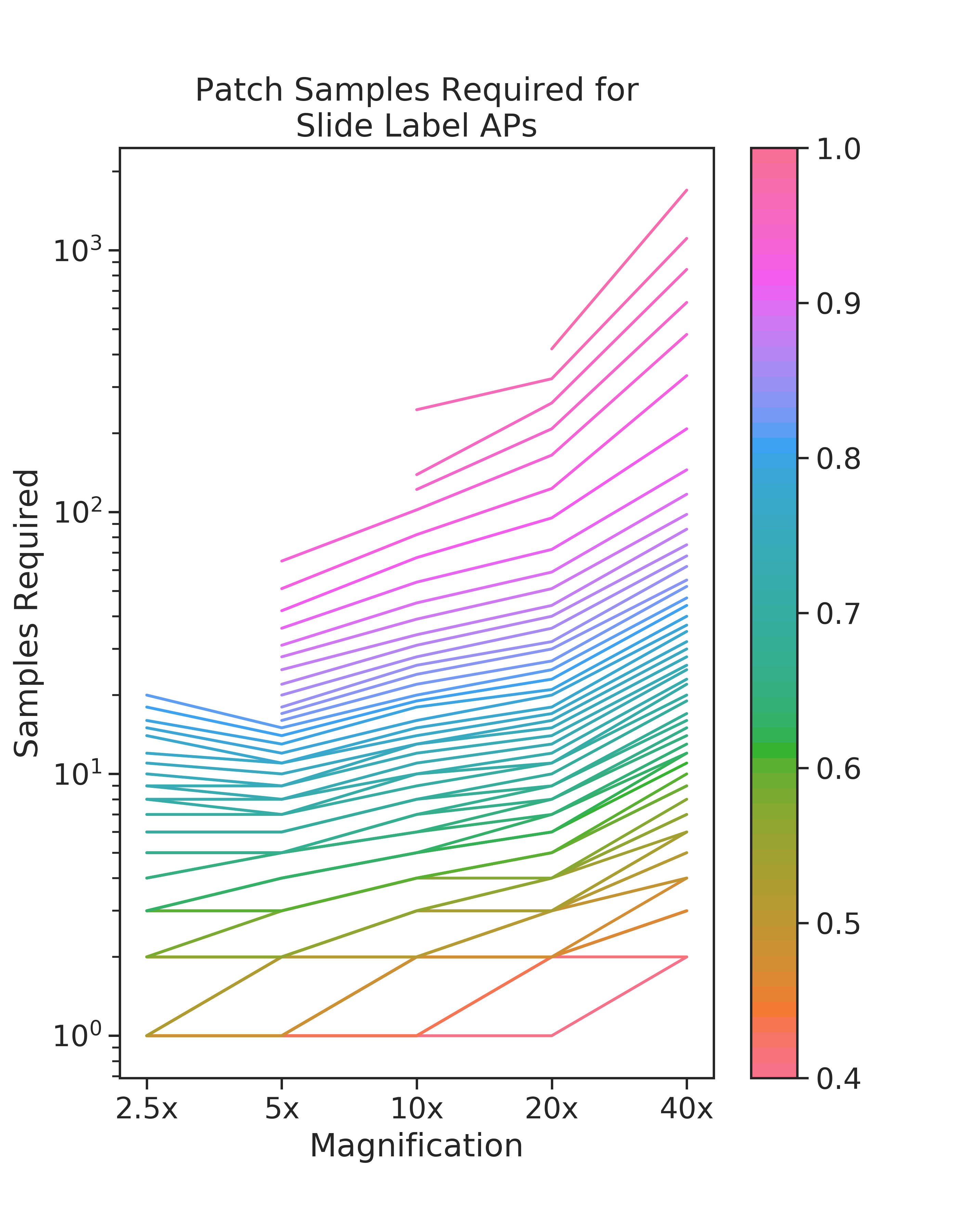}
    \caption{Number of patches sampled vs. magnification vs. AP of slide-level classification for all slides in the test set at varying slide magnifications. AP is presented as a function of the number of patches sampled, averaged over 500 random patch orderings. The cost is represented by the number of patches processed, and is proportional to the processing time, when each patch is processed in sequence.}
    \label{fig:rainbow_mag}
\end{figure}

\begingroup
\begin{table*}
\caption {\label{tab:table2} Slide classification  metrics as average precision (AP) and area under the receiver operating characteristic curve (AUC), with times reported as ranges of the number of CNN evaluations for which the performance threshold is within one standard deviation of the mean value of the classification metric over 500 uniformly random patch orderings.} 
\centering
\begin{tabular}{llllllllllll}
& \multicolumn{5}{c}{Patch Samples Required for .95 ($\pm$ SD) AP} & & \multicolumn{5}{c}{Patch Samples Required for .98 ($\pm$ SD) AUC}
\\
\cline{2-6} \cline{8-12}
\\
 \multicolumn{1}{c}{Training Set Size} & \multicolumn{5}{c}{Patch Magnification} & & \multicolumn{5}{c}{Patch Magnification}\\
\cline{2-6} \cline{8-12}
&  2.5x  & 5x & 10x & 20x & 40x &
&  2.5x  & 5x & 10x & 20x & 40x \\
\cline{2-6} \cline{8-12}
8   & - & - & - & - & - &  & - & - & - & - & -\\
16   & - & - & 123-287 & 186-450 & 557-1847 &
& - & - & 100-232 & 179-460 & 481-1749\\
24   & -  & 71- & 60-147 & 96-292 & 219-906 &
& - & 47-77 & 57-133 & 104-321 & 201-901\\
32   & -  & 67- & 79-162 & 114-308 & 294-1020 &
& - & 43- & 81-164 & 117-353 & 324-1259 

\end{tabular}
\end{table*}
\endgroup

Examining first the AP as a function of patches sampled when the full training and validation sets were used, we observed that the magnification of the model with the highest AP increased monotonically with the number of patches sampled, from 2.5x to 20x, with the 40x model never becoming the dominant model (see Figure \ref{fig:cost_vs_AP}). 

Higher classification performance required a greater number of patch samples, with the cost to reach a given performance threshold dependent on the patch magnification. We quantified how the magnification affects the cost required to reach mean APs between .40 and 1.00 in Figure \ref{fig:rainbow_mag}. We found that the maximum AP achievable by each model increased with the magnification, and that, among the models achieving a given AP, the lowest magnification model required far fewer patch samples than the other models. For APs up to approximately 0.68, the 2.5x model performed slide classification most efficiently, with under 10 patches having to be sampled on average. For APs of 0.63 to roughly 0.96, the 5x model was most efficient, reaching a 0.96 AP with fewer than 90 patches having to be sampled on average. For APs between roughly 0.96 and 0.99, the 10x model dominated, requiring fewer than 250 patches to be sampled. For APs between 0.99 and 1.00, the 20x model was the most efficient, requiring at least 400 patches to be sampled, in order to reach a perfect AP of 1.00.
As the 20x model saturated at an AP of 1.00, the 40x model was never the most efficient, requiring over 100 patches to reach this threshold. Similar trends held for the AUC metric, with higher values for AUC than AP at each number of patches sampled, and all magnifications above 2.5x able to reach AUCs above 0.95. For APs between 0.97 and 0.99, the 10x model required the fewest samples, and is a promising choice of patch magnification for clinical settings, if computational hardware constraints are present which might impact inference time, with the 20x model being preferred otherwise.

\begin{figure}[ht!]
    \includegraphics[width=\columnwidth]{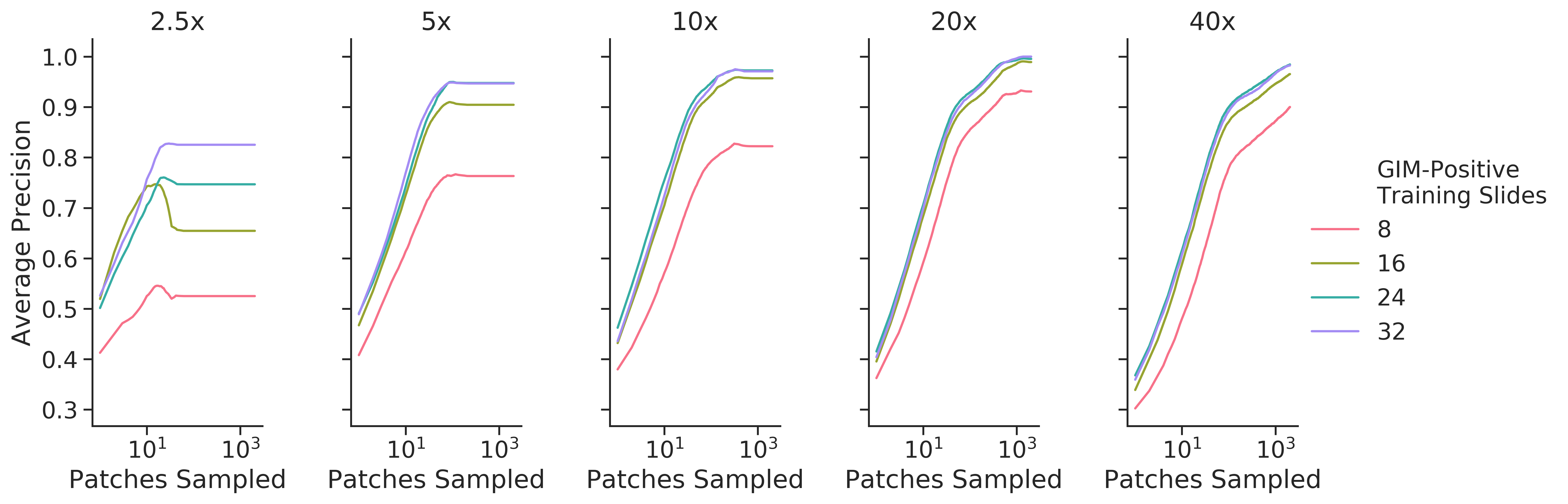}
    \caption{Number of patches sampled vs. Average Precision of slide-level classification for all GIM-positive slides in the test set, for varying training set sizes (numbers of annotated positive slides), across all magnifications. Classifiers were trained at 5 patch magnification levels, with varying quantities of GIM-positive slides in the training set.}
    \label{fig:cost_vs_training_size}
\end{figure}

Examining the tradeoff between AP and number of patches sampled for each supervision group in Figure \ref{fig:cost_vs_training_size}, we found that increasing the size of the training and validation sets led to the greatest benefit at lower magnifications (2.5x and 5x). At higher magnifications, the difference in the curves for AP versus number of patches sampled was negligible for the 3 groups with the largest training and validation sets. Intuitively, this trend can be explained by the larger effective tissue area contained within lower versus higher magnification patches, which allows for greater histomorphologic heterogeneity in tissue architectural features, necessitating larger training and validation sets for improved model generalizability. (Due to the nature of histopathologic specimens, higher magnification patches generally contain more uniformity in the types and arrangements of cells and other morphologic features.)

Our results indicate that lower magnifications can be more efficient, if moderate performance is required. For higher performance, higher magnifications are needed, though at a higher cost. Patches at 40x are too numerous for the 40x model to dominate over any of the other magnifications, using our inference strategy.

\subsection{Inference Times}
While the numbers of patch samples required to perform evidence localization and slide classification are useful hardware-agnostic measures for the cost of each model at inference time, we additionally measured inference times when models are run in a web browser using consumer-grade workstation hardware that reasonably approximates the computational resources available to the average pathologist. In particular, we measured the time required to process a single patch using Tensorflow.js \cite{smilkov2019tensorflowjs} on a 2015 MacBook Pro with a 2.8 GHz Intel Core i7 CPU and Intel Iris Pro GPU in Google Chrome Version 84. Inference in Tensorflow.js may be executed using any of three backends, with the choice of backend depending on web browser support and available hardware. Specifically, it can be performed using either vanilla Javascript on a CPU (``cpu''), WebAssembly on a CPU (``wasm''), or WebGL on a graphics card (``webgl''). The ``cpu'' backend is slowest, but has the lowest hardware and software requirements, whereas the ``wasm'' backend requires browser support for WebAssembly, and the ``webgl'' backend requires browser support for WebGL, in addition to a graphics card. The mean time required to perform inference on a single patch, averaged over 100 patch evaluations using the CNN, was 3 seconds for the ``cpu'' backend, 0.28 seconds for the ``wasm'' backend, and 0.05 seconds for the ``webgl'' backend.

We found that, even in the absence of a GPU, the random patch sampling method could perform evidence localization and slide classification in under one minute using the ``wasm'' backend, which is well within desired clinical time constraints. The 10x model was able to perform evidence localization with a zero failure rate, and Table II shows that it was able to perform slide classification with an AUC of 0.98 within a range of 81-164 patch samples, corresponding to roughly 4-8 minutes per WSI with the ``cpu'' backend, 22-45 seconds with the ``wasm'' backend, and 4-8 seconds with the ``webgl'' backend. At 20x magnification, maximum performance metrics were attained within 500 patch samples, corresponding to inference times of 25 seconds with ``webgl''. In contrast, when WebGL is not available, performing inference on 500 patches with the ``cpu'' and ``wasm'' inference modes could take up to 25 minutes and 2.3 minutes respectively, placing the 20x model outside the range of near-real-time clinical utility without GPU hardware support.

\section{Discussion}

Our random patch-sampling inference strategy achieved high slide-level classification performance and high inference speeds (requiring only a fraction of the time necessary for processing an entire slide). This approach may be particularly practical in clinical settings, as it does not require the use of dedicated graphics hardware at inference time, and can easily be accessed by users in a browser interface, without sensitive patient data needing to leave the pathologist's workstation.

A limitation of our study is that we only had access to data from a single institution. It remains for future studies to investigate how well the performance characteristics of our models generalize to WSI from other institutions. Furthermore, our work does not investigate the effect of incorporating preprocessing procedures such as stain normalization, which has been shown to increase generalizability across different institutions and histology labs \cite {anghel2019high, tellez2019quantifying}. An interesting extension of our work, which might yield higher-performance classification at lower magnifications, would be to investigate the use of more fine-grained labels for patch classifier training (for example, by using goblet cell segmentation maps as labels, and aggregating a segmentation model's output into a final goblet cell score). Furthermore, investigating multi-scale models which incorporate classifications of patches at various sizes could lead to faster inference times, with our evaluation framework able to precisely quantify tradeoffs in performance and inference time for such extensions.

\section{Conclusion}
In this study, we have presented a novel WSI analysis method that sparsely samples patches from WSI of gastric biopsies for rapid diagnosis of intestinal metaplasia (GIM), which can easily be utilized by pathologists via local processing in a web browser. We have also developed an evaluation framework for quantifying tradeoffs between performance and inference time across different slide magnifications and training set sizes. We demonstrate that our random patch sampling strategy can perform slide classification with a mean AUC of greater than 0.98, by processing less than 40\% of the average slide area. We further show that this method does not require large amounts of pixel-level annotations for training, achieving high diagnostic performance with only 30 annotated GIM-positive slides. Our slide inference and evaluation methodology is inherently generalizable, and can readily be extended to other biomarker and object localization tasks in histopathology.



%




%

\bibliography{main}

\begin{thebibliography}{10}
\providecommand{\url}[1]{#1}
\csname url@samestyle\endcsname
\providecommand{\newblock}{\relax}
\providecommand{\bibinfo}[2]{#2}
\providecommand{\BIBentrySTDinterwordspacing}{\spaceskip=0pt\relax}
\providecommand{\BIBentryALTinterwordstretchfactor}{4}
\providecommand{\BIBentryALTinterwordspacing}{\spaceskip=\fontdimen2\font plus
\BIBentryALTinterwordstretchfactor\fontdimen3\font minus
  \fontdimen4\font\relax}
\providecommand{\BIBforeignlanguage}[2]{{%
\expandafter\ifx\csname l@#1\endcsname\relax
\typeout{** WARNING: IEEEtran.bst: No hyphenation pattern has been}%
\typeout{** loaded for the language `#1'. Using the pattern for}%
\typeout{** the default language instead.}%
\else
\language=\csname l@#1\endcsname
\fi
#2}}
\providecommand{\BIBdecl}{\relax}
\BIBdecl

\bibitem{dimitriou2019deep}
N.~Dimitriou, O.~Arandjelovi{\'c}, and P.~D. Caie, ``Deep learning for whole
  slide image analysis: An overview,'' \emph{Frontiers in Medicine}, vol.~6,
  2019.

\bibitem{survey}
C.~L. Srinidhi, O.~Ciga, and A.~L. Martel, ``Deep neural network models for
  computational histopathology: A survey,'' \emph{arXiv preprint
  arXiv:1912.12378}, 2019.

\bibitem{jiang2020emerging}
Y.~Jiang, M.~Yang, S.~Wang, X.~Li, and Y.~Sun, ``Emerging role of deep
  learning-based artificial intelligence in tumor pathology,'' \emph{Cancer
  Communications}, vol.~40, no.~4, pp. 154--166, 2020.

\bibitem{serag2019translational}
A.~Serag, A.~Ion-Margineanu, H.~Qureshi, R.~McMillan, M.-J. Saint~Martin,
  J.~Diamond, P.~O'Reilly, and P.~Hamilton, ``Translational ai and deep
  learning in diagnostic pathology,'' \emph{Frontiers in Medicine}, vol.~6,
  2019.

\bibitem{steiner2018impact}
D.~F. Steiner, R.~MacDonald, Y.~Liu, P.~Truszkowski, J.~D. Hipp, C.~Gammage,
  F.~Thng, L.~Peng, and M.~C. Stumpe, ``Impact of deep learning assistance on
  the histopathologic review of lymph nodes for metastatic breast cancer,''
  \emph{The American journal of surgical pathology}, vol.~42, no.~12, p. 1636,
  2018.

\bibitem{lin2019fast}
H.~Lin, H.~Chen, S.~Graham, Q.~Dou, N.~Rajpoot, and P.-A. Heng, ``Fast scannet:
  Fast and dense analysis of multi-gigapixel whole-slide images for cancer
  metastasis detection,'' \emph{IEEE transactions on medical imaging}, vol.~38,
  no.~8, pp. 1948--1958, 2019.

\bibitem{janowczyk2016deep}
A.~Janowczyk and A.~Madabhushi, ``Deep learning for digital pathology image
  analysis: A comprehensive tutorial with selected use cases,'' \emph{Journal
  of pathology informatics}, vol.~7, 2016.

\bibitem{bokhorst2018automatic}
J.-M. Bokhorst, L.~Rijstenberg, D.~Goudkade, I.~Nagtegaal, J.~van~der Laak, and
  F.~Ciompi, ``Automatic detection of tumor budding in colorectal carcinoma
  with deep learning,'' in \emph{Computational Pathology and Ophthalmic Medical
  Image Analysis}.\hskip 1em plus 0.5em minus 0.4em\relax Springer, 2018, pp.
  130--138.

\bibitem{tomita2019attention}
N.~Tomita, B.~Abdollahi, J.~Wei, B.~Ren, A.~Suriawinata, and S.~Hassanpour,
  ``Attention-based deep neural networks for detection of cancerous and
  precancerous esophagus tissue on histopathological slides,'' \emph{JAMA
  network open}, vol.~2, no.~11, pp. e1\,914\,645--e1\,914\,645, 2019.

\bibitem{jencks2018overview}
D.~S. Jencks, J.~D. Adam, M.~L. Borum, J.~M. Koh, S.~Stephen, and D.~B. Doman,
  ``Overview of current concepts in gastric intestinal metaplasia and gastric
  cancer,'' \emph{Gastroenterology \& hepatology}, vol.~14, no.~2, p.~92, 2018.

\bibitem{piazuelo2013gastric}
M.~B. Piazuelo and P.~Correa, ``Gastric cancer: overview,'' \emph{Colombia
  Medica}, vol.~44, no.~3, pp. 192--201, 2013.

\bibitem{zullo2012follow}
A.~Zullo, C.~Hassan, A.~Romiti, M.~Giusto, C.~Guerriero, R.~Lorenzetti, S.~M.
  Campo, and S.~Tomao, ``Follow-up of intestinal metaplasia in the stomach:
  When, how and why,'' \emph{World journal of gastrointestinal oncology},
  vol.~4, no.~3, p.~30, 2012.

\bibitem{paull1976histologic}
A.~Paull, J.~S. Trier, M.~D. Dalton, R.~C. Camp, P.~Loeb, and R.~K. Goyal,
  ``The histologic spectrum of barrett's esophagus,'' \emph{New England Journal
  of Medicine}, vol. 295, no.~9, pp. 476--480, 1976.

\bibitem{TGH5866}
\BIBentryALTinterwordspacing
J.~R. White and M.~Banks, ``Identifying the pre-malignant stomach: from
  guidelines to practice,'' \emph{Translational Gastroenterology and
  Hepatology}, vol.~0, no.~0, 2020. [Online]. Available:
  \url{http://tgh.amegroups.com/article/view/5866}
\BIBentrySTDinterwordspacing

\bibitem{cruz2018high}
A.~Cruz-Roa, H.~Gilmore, A.~Basavanhally, M.~Feldman, S.~Ganesan, N.~Shih,
  J.~Tomaszewski, A.~Madabhushi, and F.~Gonz{\'a}lez, ``High-throughput
  adaptive sampling for whole-slide histopathology image analysis (hashi) via
  convolutional neural networks: Application to invasive breast cancer
  detection,'' \emph{PloS one}, vol.~13, no.~5, p. e0196828, 2018.

\bibitem{huang2011time}
C.-H. Huang, A.~Veillard, L.~Roux, N.~Lom{\'e}nie, and D.~Racoceanu,
  ``Time-efficient sparse analysis of histopathological whole slide images,''
  \emph{Computerized medical imaging and graphics}, vol.~35, no. 7-8, pp.
  579--591, 2011.

\bibitem{santos2016literature}
T.~Santos and R.~Kern, ``A literature survey of early time series
  classification and deep learning.'' in \emph{Sami@ iknow}, 2016.

\bibitem{xing2012early}
Z.~Xing, J.~Pei, and S.~Y. Philip, ``Early classification on time series,''
  \emph{Knowledge and information systems}, vol.~31, no.~1, pp. 105--127, 2012.

\bibitem{dennis2018multiple}
D.~Dennis, C.~Pabbaraju, H.~V. Simhadri, and P.~Jain, ``Multiple instance
  learning for efficient sequential data classification on resource-constrained
  devices,'' in \emph{Advances in Neural Information Processing Systems}, 2018,
  pp. 10\,953--10\,964.

\bibitem{bankhead2017qupath}
P.~Bankhead, M.~B. Loughrey, J.~A. Fern{\'a}ndez, Y.~Dombrowski, D.~G. McArt,
  P.~D. Dunne, S.~McQuaid, R.~T. Gray, L.~J. Murray, H.~G. Coleman
  \emph{et~al.}, ``Qupath: Open source software for digital pathology image
  analysis,'' \emph{Scientific reports}, vol.~7, no.~1, pp. 1--7, 2017.

\bibitem{ehsan2015integral}
S.~Ehsan, A.~F. Clark, N.~U. Rehman, and K.~D. McDonald-Maier, ``Integral
  images: Efficient algorithms for their computation and storage in
  resource-constrained embedded vision systems,'' \emph{Sensors}, vol.~15,
  no.~7, pp. 16\,804--16\,830, 2015.

\bibitem{otsu1979threshold}
N.~Otsu, ``A threshold selection method from gray-level histograms,''
  \emph{IEEE transactions on systems, man, and cybernetics}, vol.~9, no.~1, pp.
  62--66, 1979.

\bibitem{coudray_classification_2018}
\BIBentryALTinterwordspacing
N.~Coudray, P.~S. Ocampo, T.~Sakellaropoulos, N.~Narula, M.~Snuderl, D.~Fenyö,
  A.~L. Moreira, N.~Razavian, and A.~Tsirigos,
  ``\BIBforeignlanguage{en}{Classification and mutation prediction from
  non–small cell lung cancer histopathology images using deep learning},''
  \emph{\BIBforeignlanguage{en}{Nature Medicine}}, vol.~24, no.~10, pp.
  1559--1567, Oct. 2018. [Online]. Available:
  \url{https://www.nature.com/articles/s41591-018-0177-5}
\BIBentrySTDinterwordspacing

\bibitem{kraus2016classifying}
O.~Z. Kraus, J.~L. Ba, and B.~J. Frey, ``Classifying and segmenting microscopy
  images with deep multiple instance learning,'' \emph{Bioinformatics},
  vol.~32, no.~12, pp. i52--i59, 2016.

\bibitem{tan_efficientnet_2019}
\BIBentryALTinterwordspacing
M.~Tan and Q.~V. Le, ``{EfficientNet}: {Rethinking} {Model} {Scaling} for
  {Convolutional} {Neural} {Networks},'' \emph{arXiv:1905.11946 [cs, stat]},
  Nov. 2019, arXiv: 1905.11946. [Online]. Available:
  \url{http://arxiv.org/abs/1905.11946}
\BIBentrySTDinterwordspacing

\bibitem{he2016deep}
K.~He, X.~Zhang, S.~Ren, and J.~Sun, ``Deep residual learning for image
  recognition,'' in \emph{Proceedings of the IEEE conference on computer vision
  and pattern recognition}, 2016, pp. 770--778.

\bibitem{deng2009imagenet}
J.~Deng, W.~Dong, R.~Socher, L.-J. Li, K.~Li, and L.~Fei-Fei, ``Imagenet: A
  large-scale hierarchical image database,'' in \emph{2009 IEEE conference on
  computer vision and pattern recognition}.\hskip 1em plus 0.5em minus
  0.4em\relax Ieee, 2009, pp. 248--255.

\bibitem{JMLR:v18:16-365}
\BIBentryALTinterwordspacing
G.~Lema{{\^i}}tre, F.~Nogueira, and C.~K. Aridas, ``Imbalanced-learn: A python
  toolbox to tackle the curse of imbalanced datasets in machine learning,''
  \emph{Journal of Machine Learning Research}, vol.~18, no.~17, pp. 1--5, 2017.
  [Online]. Available: \url{http://jmlr.org/papers/v18/16-365}
\BIBentrySTDinterwordspacing

\bibitem{kingma2014adam}
D.~P. Kingma and J.~Ba, ``Adam: A method for stochastic optimization,''
  \emph{arXiv preprint arXiv:1412.6980}, 2014.

\bibitem{smilkov2019tensorflowjs}
D.~Smilkov, N.~Thorat, Y.~Assogba, A.~Yuan, N.~Kreeger, P.~Yu, K.~Zhang,
  S.~Cai, E.~Nielsen, D.~Soergel, S.~Bileschi, M.~Terry, C.~Nicholson, S.~N.
  Gupta, S.~Sirajuddin, D.~Sculley, R.~Monga, G.~Corrado, F.~B. Viégas, and
  M.~Wattenberg, ``Tensorflow.js: Machine learning for the web and beyond,''
  2019.

\bibitem{anghel2019high}
A.~Anghel, M.~Stanisavljevic, S.~Andani, N.~Papandreou, J.~H. R{\"u}schoff,
  P.~Wild, M.~Gabrani, and H.~Pozidis, ``A high-performance system for robust
  stain normalization of whole-slide images in histopathology,''
  \emph{Frontiers in Medicine}, vol.~6, 2019.

\bibitem{tellez2019quantifying}
D.~Tellez, G.~Litjens, P.~B{\'a}ndi, W.~Bulten, J.-M. Bokhorst, F.~Ciompi, and
  J.~van~der Laak, ``Quantifying the effects of data augmentation and stain
  color normalization in convolutional neural networks for computational
  pathology,'' \emph{Medical image analysis}, vol.~58, p. 101544, 2019.

\end{thebibliography}
\bibliographystyle{IEEEtran}

@article{jencks2018overview,
  title={Overview of current concepts in gastric intestinal metaplasia and gastric cancer},
  author={Jencks, David S and Adam, Jason D and Borum, Marie L and Koh, Joyce M and Stephen, Sindu and Doman, David B},
  journal={Gastroenterology \& hepatology},
  volume={14},
  number={2},
  pages={92},
  year={2018},
  publisher={Millenium Medical Publishing}
}
@article{TGH5866,
	author = {Jonathan R. White and Matthew Banks},
	title = {Identifying the pre-malignant stomach: from guidelines to practice},
	journal = {Translational Gastroenterology and Hepatology},
	volume = {0},
	number = {0},
	year = {2020},
	keywords = {},
	abstract = {},
	url = {http://tgh.amegroups.com/article/view/5866}
}
@article{zullo2012follow,
  title={Follow-up of intestinal metaplasia in the stomach: When, how and why},
  author={Zullo, Angelo and Hassan, Cesare and Romiti, Adriana and Giusto, Michela and Guerriero, Carmine and Lorenzetti, Roberto and Campo, Salvatore MA and Tomao, Silverio},
  journal={World journal of gastrointestinal oncology},
  volume={4},
  number={3},
  pages={30},
  year={2012},
  publisher={Baishideng Publishing Group Inc}
}

@article{paull1976histologic,
  title={The histologic spectrum of Barrett's esophagus},
  author={Paull, Andrew and Trier, Jerry S and Dalton, M David and Camp, Roger C and Loeb, Peter and Goyal, Raj K},
  journal={New England Journal of Medicine},
  volume={295},
  number={9},
  pages={476--480},
  year={1976},
  publisher={Mass Medical Soc}
}

@article{correa2010pathology,
  title={Pathology of gastric intestinal metaplasia: clinical implications},
  author={Correa, Pelayo and Piazuelo, M Blanca and Wilson, Keith T},
  journal={The American journal of gastroenterology},
  volume={105},
  number={3},
  pages={493},
  year={2010},
  publisher={NIH Public Access}
}

@article{coudray_classification_2018,
	title = {Classification and mutation prediction from non–small cell lung cancer histopathology images using deep learning},
	volume = {24},
	copyright = {2018 The Author(s), under exclusive licence to Springer Nature America, Inc.},
	issn = {1546-170X},
	url = {https://www.nature.com/articles/s41591-018-0177-5},
	doi = {10.1038/s41591-018-0177-5},
	abstract = {A convolutional neural network model using feature extraction and machine-learning techniques provides a tool for classification of lung cancer histopathology images and predicting mutational status of driver oncogenes},
	language = {en},
	number = {10},
	urldate = {2020-01-27},
	journal = {Nature Medicine},
	author = {Coudray, Nicolas and Ocampo, Paolo Santiago and Sakellaropoulos, Theodore and Narula, Navneet and Snuderl, Matija and Fenyö, David and Moreira, Andre L. and Razavian, Narges and Tsirigos, Aristotelis},
	month = oct,
	year = {2018},
	pages = {1559--1567},
	file = {Full Text PDF:/Users/rajpurkar/Zotero/storage/MVEFY6NW/Coudray et al. - 2018 - Classification and mutation prediction from non–sm.pdf:application/pdf;Snapshot:/Users/rajpurkar/Zotero/storage/SZIT7ALS/s41591-018-0177-5.html:text/html}
}

@article{dimitriou2019deep,
  title={Deep learning for whole slide image analysis: An overview},
  author={Dimitriou, Neofytos and Arandjelovi{\'c}, Ognjen and Caie, Peter D},
  journal={Frontiers in Medicine},
  volume={6},
  year={2019},
  publisher={Frontiers Media SA}
}

@article{kather_deep_2019,
	title = {Deep learning can predict microsatellite instability directly from histology in gastrointestinal cancer},
	volume = {25},
	copyright = {2019 The Author(s), under exclusive licence to Springer Nature America, Inc.},
	issn = {1546-170X},
	url = {https://www.nature.com/articles/s41591-019-0462-y},
	doi = {10.1038/s41591-019-0462-y},
	abstract = {A deep residual learning framework identifies microsatellite instability in histology slides from patients with cancer and can be used to guide immunotherapy.},
	language = {en},
	number = {7},
	urldate = {2020-01-27},
	journal = {Nature Medicine},
	author = {Kather, Jakob Nikolas and Pearson, Alexander T. and Halama, Niels and Jäger, Dirk and Krause, Jeremias and Loosen, Sven H. and Marx, Alexander and Boor, Peter and Tacke, Frank and Neumann, Ulf Peter and Grabsch, Heike I. and Yoshikawa, Takaki and Brenner, Hermann and Chang-Claude, Jenny and Hoffmeister, Michael and Trautwein, Christian and Luedde, Tom},
	month = jul,
	year = {2019},
	pages = {1054--1056},
	file = {Full Text PDF:/Users/rajpurkar/Zotero/storage/R6BPEI8W/Kather et al. - 2019 - Deep learning can predict microsatellite instabili.pdf:application/pdf}
}

@article{steiner2018impact,
  title={Impact of deep learning assistance on the histopathologic review of lymph nodes for metastatic breast cancer},
  author={Steiner, David F and MacDonald, Robert and Liu, Yun and Truszkowski, Peter and Hipp, Jason D and Gammage, Christopher and Thng, Florence and Peng, Lily and Stumpe, Martin C},
  journal={The American journal of surgical pathology},
  volume={42},
  number={12},
  pages={1636},
  year={2018},
  publisher={Wolters Kluwer Health}
}

@article{tizhoosh_artificial_2018,
	title = {Artificial {Intelligence} and {Digital} {Pathology}: {Challenges} and {Opportunities}},
	volume = {9},
	issn = {2229-5089},
	shorttitle = {Artificial {Intelligence} and {Digital} {Pathology}},
	url = {https://www.ncbi.nlm.nih.gov/pmc/articles/PMC6289004/},
	doi = {10.4103/jpi.jpi_53_18},
	abstract = {In light of the recent success of artificial intelligence (AI) in computer vision applications, many researchers and physicians expect that AI would be able to assist in many tasks in digital pathology. Although opportunities are both manifest and tangible, there are clearly many challenges that need to be overcome in order to exploit the AI potentials in computational pathology. In this paper, we strive to provide a realistic account of all challenges and opportunities of adopting AI algorithms in digital pathology from both engineering and pathology perspectives.},
	urldate = {2020-01-26},
	journal = {Journal of Pathology Informatics},
	author = {Tizhoosh, Hamid Reza and Pantanowitz, Liron},
	month = nov,
	year = {2018},
	pmid = {30607305},
	pmcid = {PMC6289004},
	file = {Submitted Version:/Users/rajpurkar/Zotero/storage/KH54MFBG/Tizhoosh and Pantanowitz - 2018 - Artificial Intelligence and Digital Pathology Cha.pdf:application/pdf}
}

@article{tan_efficientnet_2019,
	title = {{EfficientNet}: {Rethinking} {Model} {Scaling} for {Convolutional} {Neural} {Networks}},
	shorttitle = {{EfficientNet}},
	url = {http://arxiv.org/abs/1905.11946},
	abstract = {Convolutional Neural Networks (ConvNets) are commonly developed at a fixed resource budget, and then scaled up for better accuracy if more resources are available. In this paper, we systematically study model scaling and identify that carefully balancing network depth, width, and resolution can lead to better performance. Based on this observation, we propose a new scaling method that uniformly scales all dimensions of depth/width/resolution using a simple yet highly effective compound coefficient. We demonstrate the effectiveness of this method on scaling up MobileNets and ResNet. To go even further, we use neural architecture search to design a new baseline network and scale it up to obtain a family of models, called EfficientNets, which achieve much better accuracy and efficiency than previous ConvNets. In particular, our EfficientNet-B7 achieves state-of-the-art 84.4\% top-1 / 97.1\% top-5 accuracy on ImageNet, while being 8.4x smaller and 6.1x faster on inference than the best existing ConvNet. Our EfficientNets also transfer well and achieve state-of-the-art accuracy on CIFAR-100 (91.7\%), Flowers (98.8\%), and 3 other transfer learning datasets, with an order of magnitude fewer parameters. Source code is at https://github.com/tensorflow/tpu/tree/master/models/official/efficientnet.},
	urldate = {2020-01-27},
	journal = {arXiv:1905.11946 [cs, stat]},
	author = {Tan, Mingxing and Le, Quoc V.},
	month = nov,
	year = {2019},
	note = {arXiv: 1905.11946},
	keywords = {Computer Science - Computer Vision and Pattern Recognition, Computer Science - Machine Learning, Statistics - Machine Learning},
	file = {arXiv Fulltext PDF:/Users/rajpurkar/Zotero/storage/R62IKS3P/Tan and Le - 2019 - EfficientNet Rethinking Model Scaling for Convolu.pdf:application/pdf;arXiv.org Snapshot:/Users/rajpurkar/Zotero/storage/E8TQ2RDW/1905.html:text/html}
}

@article{janowczyk2016deep,
  title={Deep learning for digital pathology image analysis: A comprehensive tutorial with selected use cases},
  author={Janowczyk, Andrew and Madabhushi, Anant},
  journal={Journal of pathology informatics},
  volume={7},
  year={2016},
  publisher={Wolters Kluwer--Medknow Publications}
}

@incollection{bokhorst2018automatic,
  title={Automatic Detection of Tumor Budding in Colorectal Carcinoma with Deep Learning},
  author={Bokhorst, John-Melle and Rijstenberg, Lucia and Goudkade, Danny and Nagtegaal, Iris and van der Laak, Jeroen and Ciompi, Francesco},
  booktitle={Computational Pathology and Ophthalmic Medical Image Analysis},
  pages={130--138},
  year={2018},
  publisher={Springer}
}

@article{JMLR:v18:16-365,
author  = {Guillaume  Lema{{\^i}}tre and Fernando Nogueira and Christos K. Aridas},
title   = {Imbalanced-learn: A Python Toolbox to Tackle the Curse of Imbalanced Datasets in Machine Learning},
journal = {Journal of Machine Learning Research},
year    = {2017},
volume  = {18},
number  = {17},
pages   = {1-5},
url     = {http://jmlr.org/papers/v18/16-365}
}

@article{yu_predicting_2016,
    title = {Predicting non-small cell lung cancer prognosis by fully automated microscopic pathology image features},
    volume = {7},
    copyright = {2016 The Author(s)},
    issn = {2041-1723},
    url = {https://www.nature.com/articles/ncomms12474},
    doi = {10.1038/ncomms12474},
    abstract = {Diagnosis of lung cancer through manual histopathology evaluation is insufficient to predict patient survival. Here, the authors use computerized image processing to identify diagnostically relevant image features and use these features to distinguish lung cancer patients with different prognoses.},
    language = {en},
    number = {1},
    urldate = {2020-01-27},
    journal = {Nature Communications},
    author = {Yu, Kun-Hsing and Zhang, Ce and Berry, Gerald J. and Altman, Russ B. and Ré, Christopher and Rubin, Daniel L. and Snyder, Michael},
    month = aug,
    year = {2016},
    pages = {1--10},
    file = {Full Text PDF:/Users/rajpurkar/Zotero/storage/VWYXUZN8/Yu et al. - 2016 - Predicting non-small cell lung cancer prognosis by.pdf:application/pdf}
}
@article{campanella_clinical-grade_2019,
    title = {Clinical-grade computational pathology using weakly supervised deep learning on whole slide images},
    volume = {25},
    copyright = {2019 The Author(s), under exclusive licence to Springer Nature America, Inc.},
    issn = {1546-170X},
    url = {https://www.nature.com/articles/s41591-019-0508-1},
    doi = {10.1038/s41591-019-0508-1},
    abstract = {A deep learning model trained on real-world digital pathology data achieves clinical performance in cancer diagnosis.},
    language = {en},
    number = {8},
    urldate = {2020-01-27},
    journal = {Nature Medicine},
    author = {Campanella, Gabriele and Hanna, Matthew G. and Geneslaw, Luke and Miraflor, Allen and Werneck Krauss Silva, Vitor and Busam, Klaus J. and Brogi, Edi and Reuter, Victor E. and Klimstra, David S. and Fuchs, Thomas J.},
    month = aug,
    year = {2019},
    pages = {1301--1309},
    file = {Full Text PDF:/Users/rajpurkar/Zotero/storage/G5WK9LTR/Campanella et al. - 2019 - Clinical-grade computational pathology using weakl.pdf:application/pdf;Snapshot:/Users/rajpurkar/Zotero/storage/A8PUYFNC/s41591-019-0508-1.html:text/html}
}
@article{qaiser_learning_2019,
    title = {Learning {Where} to {See}: {A} {Novel} {Attention} {Model} for {Automated} {Immunohistochemical} {Scoring}},
    volume = {38},
    issn = {1558-254X},
    shorttitle = {Learning {Where} to {See}},
    doi = {10.1109/TMI.2019.2907049},
    abstract = {Estimating over-amplification of human epidermal growth factor receptor 2 (HER2) on invasive breast cancer is regarded as a significant predictive and prognostic marker. We propose a novel deep reinforcement learning (DRL)-based model that treats immunohistochemical (IHC) scoring of HER2 as a sequential learning task. For a given image tile sampled from multi-resolution giga-pixel whole slide image (WSI), the model learns to sequentially identify some of the diagnostically relevant regions of interest (ROIs) by following a parameterized policy. The selected ROIs are processed by recurrent and residual convolution networks to learn the discriminative features for different HER2 scores and predict the next location, without requiring to process all the sub-image patches of a given tile for predicting the HER2 score, mimicking the histopathologist who would not usually analyze every part of the slide at the highest magnification. The proposed model incorporates a task-specific regularization term and inhibition of return mechanism to prevent the model from revisiting the previously attended location.}
}
    
  @article{tomita2019attention,
  title={Attention-based deep neural networks for detection of cancerous and precancerous esophagus tissue on histopathological slides},
  author={Tomita, Naofumi and Abdollahi, Behnaz and Wei, Jason and Ren, Bing and Suriawinata, Arief and Hassanpour, Saeed},
  journal={JAMA network open},
  volume={2},
  number={11},
  pages={e1914645--e1914645},
  year={2019},
  publisher={American Medical Association}
}

@article{momeni2018deep,
  title={Deep recurrent attention models for histopathological image analysis},
  author={Momeni, Alexandre and Thibault, Marc and Gevaert, Olivier},
  journal={BioRxiv},
  pages={438341},
  year={2018},
  publisher={Cold Spring Harbor Laboratory}
}

@article{ferlay2015cancer,
  title={Cancer incidence and mortality worldwide: sources, methods and major patterns in GLOBOCAN 2012},
  author={Ferlay, Jacques and Soerjomataram, Isabelle and Dikshit, Rajesh and Eser, Sultan and Mathers, Colin and Rebelo, Marise and Parkin, Donald Maxwell and Forman, David and Bray, Freddie},
  journal={International journal of cancer},
  volume={136},
  number={5},
  pages={E359--E386},
  year={2015},
  publisher={Wiley Online Library}
}

@article{shichijo2016histologic,
  title={Histologic intestinal metaplasia and endoscopic atrophy are predictors of gastric cancer development after Helicobacter pylori eradication},
  author={Shichijo, Satoki and Hirata, Yoshihiro and Niikura, Ryota and Hayakawa, Yoku and Yamada, Atsuo and Ushiku, Tetsuo and Fukayama, Masashi and Koike, Kazuhiko},
  journal={Gastrointestinal endoscopy},
  volume={84},
  number={4},
  pages={618--624},
  year={2016},
  publisher={Elsevier}
}

@article{emura2019early,
  title={Early gastric cancer: current limitations and what can be done to address them},
  author={Emura, Fabian and Rodriguez-Reyes, Carlos and Giraldo-Cadavid, Luis},
  journal={American Journal of Gastroenterology},
  volume={114},
  number={6},
  pages={841--845},
  year={2019},
  publisher={LWW}
}

@article{piazuelo2013gastric,
  title={Gastric cancer: overview},
  author={Piazuelo, Mar{\'\i}a Blanca and Correa, Pelayo},
  journal={Colombia Medica},
  volume={44},
  number={3},
  pages={192--201},
  year={2013},
  publisher={Universidad del Valle}
}

@article{chen2020deep,
  title={Deep Learning on Computational-Resource-Limited Platforms: A Survey},
  author={Chen, Chunlei and Zhang, Peng and Zhang, Huixiang and Dai, Jiangyan and Yi, Yugen and Zhang, Huihui and Zhang, Yonghui},
  journal={Mobile Information Systems},
  volume={2020},
  year={2020},
  publisher={Hindawi}
}

@article{survey,
  title={Deep neural network models for computational histopathology: A survey},
  author={Srinidhi, Chetan L and Ciga, Ozan and Martel, Anne L},
  journal={arXiv preprint arXiv:1912.12378},
  year={2019}
}

@article{iizuka2020deep,
  title={Deep Learning Models for Histopathological Classification of Gastric and colonic epithelial tumours},
  author={Iizuka, Osamu and Kanavati, Fahdi and Kato, Kei and Rambeau, Michael and Arihiro, Koji and Tsuneki, Masayuki},
  journal={Scientific Reports},
  volume={10},
  number={1},
  pages={1--11},
  year={2020},
  publisher={Nature Publishing Group}
}

@article{jiang2020emerging,
  title={Emerging role of deep learning-based artificial intelligence in tumor pathology},
  author={Jiang, Yahui and Yang, Meng and Wang, Shuhao and Li, Xiangchun and Sun, Yan},
  journal={Cancer Communications},
  volume={40},
  number={4},
  pages={154--166},
  year={2020},
  publisher={Wiley Online Library}
}

@article{serag2019translational,
  title={Translational AI and deep learning in diagnostic pathology},
  author={Serag, Ahmed and Ion-Margineanu, Adrian and Qureshi, Hammad and McMillan, Ryan and Saint Martin, Marie-Judith and Diamond, Jim and O'Reilly, Paul and Hamilton, Peter},
  journal={Frontiers in Medicine},
  volume={6},
  year={2019},
  publisher={Frontiers Media SA}
}

@inproceedings{santos2016literature,
  title={A Literature Survey of Early Time Series Classification and Deep Learning.},
  author={Santos, Tiago and Kern, Roman},
  booktitle={Sami@ iknow},
  year={2016}
}

@article{huang2011time,
  title={Time-efficient sparse analysis of histopathological whole slide images},
  author={Huang, Chao-Hui and Veillard, Antoine and Roux, Ludovic and Lom{\'e}nie, Nicolas and Racoceanu, Daniel},
  journal={Computerized medical imaging and graphics},
  volume={35},
  number={7-8},
  pages={579--591},
  year={2011},
  publisher={Elsevier}
}

@article{xing2012early,
  title={Early classification on time series},
  author={Xing, Zhengzheng and Pei, Jian and Philip, S Yu},
  journal={Knowledge and information systems},
  volume={31},
  number={1},
  pages={105--127},
  year={2012},
  publisher={Springer}
}

@misc{smilkov2019tensorflowjs,
      title={TensorFlow.js: Machine Learning for the Web and Beyond}, 
      author={Daniel Smilkov and Nikhil Thorat and Yannick Assogba and Ann Yuan and Nick Kreeger and Ping Yu and Kangyi Zhang and Shanqing Cai and Eric Nielsen and David Soergel and Stan Bileschi and Michael Terry and Charles Nicholson and Sandeep N. Gupta and Sarah Sirajuddin and D. Sculley and Rajat Monga and Greg Corrado and Fernanda B. Viégas and Martin Wattenberg},
      year={2019},
      eprint={1901.05350},
      archivePrefix={arXiv},
      primaryClass={cs.LG}
}
@inproceedings{dennis2018multiple,
  title={Multiple instance learning for efficient sequential data classification on resource-constrained devices},
  author={Dennis, Don and Pabbaraju, Chirag and Simhadri, Harsha Vardhan and Jain, Prateek},
  booktitle={Advances in Neural Information Processing Systems},
  pages={10953--10964},
  year={2018}
}

@article{wang2019rmdl,
  title={RMDL: Recalibrated multi-instance deep learning for whole slide gastric image classification},
  author={Wang, Shujun and Zhu, Yaxi and Yu, Lequan and Chen, Hao and Lin, Huangjing and Wan, Xiangbo and Fan, Xinjuan and Heng, Pheng-Ann},
  journal={Medical image analysis},
  volume={58},
  pages={101549},
  year={2019},
  publisher={Elsevier}
}

@article{kraus2016classifying,
  title={Classifying and segmenting microscopy images with deep multiple instance learning},
  author={Kraus, Oren Z and Ba, Jimmy Lei and Frey, Brendan J},
  journal={Bioinformatics},
  volume={32},
  number={12},
  pages={i52--i59},
  year={2016},
  publisher={Oxford University Press}
}
@inproceedings{he2016deep,
  title={Deep residual learning for image recognition},
  author={He, Kaiming and Zhang, Xiangyu and Ren, Shaoqing and Sun, Jian},
  booktitle={Proceedings of the IEEE conference on computer vision and pattern recognition},
  pages={770--778},
  year={2016}
}

@inproceedings{deng2009imagenet,
  title={Imagenet: A large-scale hierarchical image database},
  author={Deng, Jia and Dong, Wei and Socher, Richard and Li, Li-Jia and Li, Kai and Fei-Fei, Li},
  booktitle={2009 IEEE conference on computer vision and pattern recognition},
  pages={248--255},
  year={2009},
  organization={Ieee}
}

@article{kingma2014adam,
  title={Adam: A method for stochastic optimization},
  author={Kingma, Diederik P and Ba, Jimmy},
  journal={arXiv preprint arXiv:1412.6980},
  year={2014}
}

@article{bankhead2017qupath,
  title={QuPath: Open source software for digital pathology image analysis},
  author={Bankhead, Peter and Loughrey, Maurice B and Fern{\'a}ndez, Jos{\'e} A and Dombrowski, Yvonne and McArt, Darragh G and Dunne, Philip D and McQuaid, Stephen and Gray, Ronan T and Murray, Liam J and Coleman, Helen G and others},
  journal={Scientific reports},
  volume={7},
  number={1},
  pages={1--7},
  year={2017},
  publisher={Nature Publishing Group}
}

@article{li2019weakly,
  title={Weakly supervised mitosis detection in breast histopathology images using concentric loss},
  author={Li, Chao and Wang, Xinggang and Liu, Wenyu and Latecki, Longin Jan and Wang, Bo and Huang, Junzhou},
  journal={Medical image analysis},
  volume={53},
  pages={165--178},
  year={2019},
  publisher={Elsevier}
}

@article{xu2014weakly,
  title={Weakly supervised histopathology cancer image segmentation and classification},
  author={Xu, Yan and Zhu, Jun-Yan and Eric, I and Chang, Chao and Lai, Maode and Tu, Zhuowen},
  journal={Medical image analysis},
  volume={18},
  number={3},
  pages={591--604},
  year={2014},
  publisher={Elsevier}
}

@inproceedings{huang2019evidence,
  title={Evidence localization for pathology images using weakly supervised learning},
  author={Huang, Yongxiang and Chung, Albert CS},
  booktitle={International Conference on Medical Image Computing and Computer-Assisted Intervention},
  pages={613--621},
  year={2019},
  organization={Springer}
}

@article{bejnordi2017context,
  title={Context-aware stacked convolutional neural networks for classification of breast carcinomas in whole-slide histopathology images},
  author={Bejnordi, Babak Ehteshami and Zuidhof, Guido and Balkenhol, Maschenka and Hermsen, Meyke and Bult, Peter and van Ginneken, Bram and Karssemeijer, Nico and Litjens, Geert and van der Laak, Jeroen},
  journal={Journal of Medical Imaging},
  volume={4},
  number={4},
  pages={044504},
  year={2017},
  publisher={International Society for Optics and Photonics}
}

@article{iizuka2020deep,
  title={Deep Learning Models for Histopathological Classification of Gastric and colonic epithelial tumours},
  author={Iizuka, Osamu and Kanavati, Fahdi and Kato, Kei and Rambeau, Michael and Arihiro, Koji and Tsuneki, Masayuki},
  journal={Scientific Reports},
  volume={10},
  number={1},
  pages={1--11},
  year={2020},
  publisher={Nature Publishing Group}
}

@inproceedings{cirecsan2013mitosis,
  title={Mitosis detection in breast cancer histology images with deep neural networks},
  author={Cire{\c{s}}an, Dan C and Giusti, Alessandro and Gambardella, Luca M and Schmidhuber, J{\"u}rgen},
  booktitle={International conference on medical image computing and computer-assisted intervention},
  pages={411--418},
  year={2013},
  organization={Springer}
}

@article{ning2019multiscale,
  title={Multiscale Context-Cascaded Ensemble Framework (MsC 2 EF): Application to Breast Histopathological Image},
  author={Ning, Zhenyuan and Zhang, Xinsen and Tu, Chao and Feng, Qianjin and Zhang, Yu},
  journal={IEEE Access},
  volume={7},
  pages={150910--150923},
  year={2019},
  publisher={IEEE}
}

@article{jayapandian2020development,
  title={Development and evaluation of deep learning--based segmentation of histologic structures in the kidney cortex with multiple histologic stains},
  author={Jayapandian, Catherine P and Chen, Yijiang and Janowczyk, Andrew R and Palmer, Matthew B and Cassol, Clarissa A and Sekulic, Miroslav and Hodgin, Jeffrey B and Zee, Jarcy and Hewitt, Stephen M and O’Toole, John and others},
  journal={Kidney International},
  year={2020},
  publisher={Elsevier}
}

@article{lin2019fast,
  title={Fast scannet: Fast and dense analysis of multi-gigapixel whole-slide images for cancer metastasis detection},
  author={Lin, Huangjing and Chen, Hao and Graham, Simon and Dou, Qi and Rajpoot, Nasir and Heng, Pheng-Ann},
  journal={IEEE transactions on medical imaging},
  volume={38},
  number={8},
  pages={1948--1958},
  year={2019},
  publisher={IEEE}
}

@article{cruz2018high,
  title={High-throughput adaptive sampling for whole-slide histopathology image analysis (HASHI) via convolutional neural networks: Application to invasive breast cancer detection},
  author={Cruz-Roa, Angel and Gilmore, Hannah and Basavanhally, Ajay and Feldman, Michael and Ganesan, Shridar and Shih, Natalie and Tomaszewski, John and Madabhushi, Anant and Gonz{\'a}lez, Fabio},
  journal={PloS one},
  volume={13},
  number={5},
  pages={e0196828},
  year={2018},
  publisher={Public Library of Science San Francisco, CA USA}
}

@inproceedings{cirecsan2013mitosis,
  title={Mitosis detection in breast cancer histology images with deep neural networks},
  author={Cire{\c{s}}an, Dan C and Giusti, Alessandro and Gambardella, Luca M and Schmidhuber, J{\"u}rgen},
  booktitle={International conference on medical image computing and computer-assisted intervention},
  pages={411--418},
  year={2013},
  organization={Springer}
}

@article{albarqouni2016aggnet,
  title={Aggnet: deep learning from crowds for mitosis detection in breast cancer histology images},
  author={Albarqouni, Shadi and Baur, Christoph and Achilles, Felix and Belagiannis, Vasileios and Demirci, Stefanie and Navab, Nassir},
  journal={IEEE transactions on medical imaging},
  volume={35},
  number={5},
  pages={1313--1321},
  year={2016},
  publisher={IEEE}
}
@article{tellez2018whole,
  title={Whole-slide mitosis detection in H\&E breast histology using PHH3 as a reference to train distilled stain-invariant convolutional networks},
  author={Tellez, David and Balkenhol, Maschenka and Otte-H{\"o}ller, Irene and van de Loo, Rob and Vogels, Rob and Bult, Peter and Wauters, Carla and Vreuls, Willem and Mol, Suzanne and Karssemeijer, Nico and others},
  journal={IEEE transactions on medical imaging},
  volume={37},
  number={9},
  pages={2126--2136},
  year={2018},
  publisher={IEEE}
}

@article{qaiser2019fast,
  title={Fast and accurate tumor segmentation of histology images using persistent homology and deep convolutional features},
  author={Qaiser, Talha and Tsang, Yee-Wah and Taniyama, Daiki and Sakamoto, Naoya and Nakane, Kazuaki and Epstein, David and Rajpoot, Nasir},
  journal={Medical image analysis},
  volume={55},
  pages={1--14},
  year={2019},
  publisher={Elsevier}
}

@article{litjens2016deep,
  title={Deep learning as a tool for increased accuracy and efficiency of histopathological diagnosis},
  author={Litjens, Geert and S{\'a}nchez, Clara I and Timofeeva, Nadya and Hermsen, Meyke and Nagtegaal, Iris and Kovacs, Iringo and Hulsbergen-Van De Kaa, Christina and Bult, Peter and Van Ginneken, Bram and Van Der Laak, Jeroen},
  journal={Scientific reports},
  volume={6},
  pages={26286},
  year={2016},
  publisher={Nature Publishing Group}
}

@article{anghel2019high,
  title={A High-Performance System for Robust Stain Normalization of Whole-Slide Images in Histopathology},
  author={Anghel, Andreea and Stanisavljevic, Milos and Andani, Sonali and Papandreou, Nikolaos and R{\"u}schoff, Jan Hendrick and Wild, Peter and Gabrani, Maria and Pozidis, Haralampos},
  journal={Frontiers in Medicine},
  volume={6},
  year={2019},
  publisher={Frontiers Media SA}
}
@article{tellez2019quantifying,
  title={Quantifying the effects of data augmentation and stain color normalization in convolutional neural networks for computational pathology},
  author={Tellez, David and Litjens, Geert and B{\'a}ndi, P{\'e}ter and Bulten, Wouter and Bokhorst, John-Melle and Ciompi, Francesco and van der Laak, Jeroen},
  journal={Medical image analysis},
  volume={58},
  pages={101544},
  year={2019},
  publisher={Elsevier}
}

@article{ehsan2015integral,
  title={Integral images: Efficient algorithms for their computation and storage in resource-constrained embedded vision systems},
  author={Ehsan, Shoaib and Clark, Adrian F and Rehman, Naveed Ur and McDonald-Maier, Klaus D},
  journal={Sensors},
  volume={15},
  number={7},
  pages={16804--16830},
  year={2015},
  publisher={Multidisciplinary Digital Publishing Institute}
}

@article{otsu1979threshold,
  title={A threshold selection method from gray-level histograms},
  author={Otsu, Nobuyuki},
  journal={IEEE transactions on systems, man, and cybernetics},
  volume={9},
  number={1},
  pages={62--66},
  year={1979},
  publisher={IEEE}
}



%








\end{document}